\documentclass[useAMS,usenatbib]{mn2e}
\usepackage{amsmath}
\usepackage{txfonts}
\usepackage{ctable,longtable}
\usepackage{graphicx}

\usepackage{footnote}
\makesavenoteenv{tabular}
\usepackage[colorlinks=true, linkcolor=blue, citecolor=blue, urlcolor=blue]{hyperref}
\title[The reliability of CII as a star formation rate indicator]{The reliability of [C{\sc{ii}}] as a star formation rate indicator}

\author[I. De Looze et al.]{
Ilse De Looze$^1$,
Maarten Baes$^1$, 
George~J. Bendo$^2$,
Luca Cortese$^3$ and
Jacopo Fritz$^1$ \\
$^1$Sterrenkundig Observatorium, Universiteit Gent, Krijgslaan 281 S9, B-9000 Gent, Belgium \\
$^2$UK ALMA Regional Centre Node, Jodrell Bank Centre for Astrophysics, School of Physics and Astronomy, \\
University of Manchester, Oxford Road, Manchester M13 9PL, United Kingdom \\
$^3$European Southern Observatory,ÊKarl-Schwarzschild Str. 2, 85748 Garching bei Muenchen, Germany 
}

\begin{document}

\date{Received 25 December 2010}

\pagerange{\pageref{firstpage}--\pageref{lastpage}} \pubyear{2010}

\maketitle

\label{firstpage}

\begin{abstract}
The [C{\sc{ii}}] 157.74$\mu$m line is an important coolant for the neutral interstellar gas. Since [C{\sc{ii}}] is the brightest spectral line for most galaxies, it is a potentially powerful tracer of star formation activity. In this paper we present a calibration of the star formation rate as a function of the [C{\sc{ii}}] luminosity for a sample of 24 star-forming galaxies in the nearby universe. This sample includes objects classified as H{\sc{ii}} regions or LINERs, but omits all Seyfert galaxies with a significant AGN contribution to the mid-infrared photometry. In order to calibrate the SFR against the line luminosity, we rely on both GALEX FUV data, which is an ideal tracer of the unobscured star formation, and MIPS 24$\mu$m, to probe the dust-enshrouded fraction of star formation. In case of normal star-forming galaxies, the [C{\sc{ii}}] luminosity correlates well with the star formation rate. However, the extension of this relation to more quiescent (H$\alpha$ EW $\le$ 10 \AA) or ultraluminous galaxies should be handled with caution, since these objects show a non-linearity in the $L_{[\text{C{\sc{ii}}}]}$-to-$L_{\text{FIR}}$ ratio as a function of $L_{\text{FIR}}$ (and thus, their star formation activity). 

We provide two possible explanations for the origin of the tight correlation between the [C{\sc{ii}}] emission and the star formation activity on a global galaxy-scale.
A first interpretation could be that the [C{\sc{ii}}] emission from PDRs arises from the immediate surroundings of star-forming regions. Since PDRs are neutral regions of warm dense gas at the boundaries between H{\sc{ii}} regions and molecular clouds and they provide the bulk of [C{\sc{ii}}] emission in most galaxies, we believe that a more or less constant contribution from these outer layers of photon-dominated molecular clumps to the [C{\sc{ii}}] emission provides a straightforward explanation for this close link between the [C{\sc{ii}}] luminosity and SFR.  
Alternatively, we consider the possibility that the [C{\sc{ii}}] emission is associated to the cold ISM, which advocates an indirect link with the star formation activity in a galaxy through the Schmidt law.
\end{abstract}

\begin{keywords}
galaxies:~star formation -- ISM: lines and bands -- infrared: galaxies -- ultraviolet: galaxies
\end{keywords}

\section{Introduction}
The [C{\sc{ii}}] line is an important coolant of the neutral interstellar medium (ISM), heated through the photoelectric effect on dust grains and Polycyclic Aromatic Hydrocarbons (PAHs), that have been exposed to ultraviolet (UV) photons. This UV emission can originate both in H{\sc{ii}} regions, the diffuse neutral and ionized interstellar medium \citep{1998ApJS..115..259N,1995ApJ...443..152W} and in PhotoDissociation Regions (PDRs) \citep{1985ApJ...291..755C,1985ApJ...291..722T,1989ApJ...344..770W,1991ApJ...377..192H,1991ApJ...373..423S,1998ApJ...499..258B}. Since the [C{\sc{ii}}] line is generally a very strong line in all star-forming galaxies and in particular for low-metallicity objects \citep{2000NewAR..44..249M}, it is a potentially powerful star formation rate (SFR) indicator. In this paper we aim to quantify the reliability of [C{\sc{ii}}] as a star formation rate indicator from a sample of galaxies in the nearby universe. 

Despite the strength of the [C{\sc{ii}}] line emission, several issues might obstruct a direct link with the star-forming activity in galaxies. For instance, the ambiguity concerning the origin of the dominant heating source for the interstellar gas, contaminates the correlation between the [C{\sc{ii}}] line emission and the SFR. Furthermore, this correlation is also affected by the saturation of the  [C{\sc{ii}}] excitation at high temperatures and high densities \citep{1999ApJ...527..795K}.
Nevertheless, using [C{\sc{ii}}] as a star formation indicator does benefit from the fact that it is unaffected by extinction in most cases. Exceptions might be extreme starbursts \citep{1998ApJ...504L..11L,2000isat.conf..337H} or edge-on galaxies \citep{1994ApJ...436..720H}. 
Recently, \citet{2010ApJ...711..757P} claimed that the deficiency of [C{\sc{ii}}] in Arp 220 is due to a high FIR/submm dust optical depth effect and is not caused by saturation of the [C{\sc{ii}}] excitation in higher density PDRs. \citet{Rangwala} (private communication) also confirms that [C{\sc{ii}}] might be affected by higher optical depths in Arp 220, but they found much lower values than \citet{2010ApJ...711..757P}. A thorough analysis of this dust obscuration effect at FIR/submm wavelengths in other galaxies with a [C{\sc{ii}}] deficiency is necessary to give insight on this debated issue. 

The Herschel satellite \citep{2010A&A...518L...1P} is currently making [C{\sc{ii}}] observations, with the aim of studying the ISM of nearby galaxies. The advent of ALMA will make this line observable at high redshifts (for redshift z$>$2, the [C{\sc{ii}}] 157.74 $\mu$m line shifts to atmospheric windows at submm/mm wavelengths, which are observable from ground-based facilities). \citet{1997ApJ...481..587S} and \citet{2006ApJ...647...60N} already probed the detectability of [C{\sc{ii}}] in high-redshift galaxies. According to \citet{2006ApJ...647...60N} very bright Lyman break galaxies and distant red galaxies are likely candidates to have the most prominent [C{\sc{ii}}] emission at high redshift. Some detections of [C{\sc{ii}}] in high-redshift objects already have been reported \citep{2005A&A...440L..51M,2006ApJ...645L..97I,2009A&A...500L...1M,2010ApJ...714L.162H,2010A&A...518L..35I,2010A&A...519L...1W,2010ApJ...724..957S}.

The [C{\sc{ii}}] line has been used previously to probe the physical processes in PDRs, but also to detect star formation in nearby galaxies \citep{2000ARA&A..38..761G,2001ApJ...561..766M,2002A&A...385..454B,2003ApJ...594..758L,2003A&A...397..871P}.
\citet{1991ApJ...373..423S} and \citet{1999MNRAS.303L..29P} even considered the use of [C{\sc{ii}}] as a diagnostic for the SFR in non-starburst galaxies. Also \citet{1999MNRAS.310..317L} found a correlation between [C{\sc{ii}}] emission and the star formation activity in an indirect way. They inferred a direct correlation of the [C{\sc{ii}}]-to-FIR flux ratio and [C{\sc{ii}}]-to-K$^\prime$-band flux ratio with the galaxy lateness, which is correlated, in its turn, with massive star formation activity. 
\citet{2002A&A...385..454B} extended this analysis and calibrated the [C{\sc{ii}}] line flux as a star formation tracer based on H$\alpha$+[N{\sc{ii}}] line fluxes. They found that the [C{\sc{ii}}] line intensity is not simply proportional to the star formation rate. Moreover, the scatter around the SFR calibration is considerable (an uncertainty of a factor $\sim$ 10 when estimating the SFR from the [C{\sc{ii}}] line emission). Considering the recent improvement in techniques to deal with [N{\sc{ii}}] contamination \citep{2007MNRAS.381..136D} and attenuation correction \citep{2009ApJ...706.1527B} of H$\alpha$ data, we will extend the work done by \citet{2002A&A...385..454B} and provide a new calibration for the SFR relation based on other reliable star formation indicators. The sample in \citet{2002A&A...385..454B} consisted of 22 nearby late-type galaxies spanning a far-IR luminosity range of $10^{8} \leq L_{\text{FIR}} \leq 10^{10.5} L_{\text{\sun}}$. We will extend this sample to more luminous galaxies up to $L_{\text{FIR}}$ $\sim$ 10$^{11.6}$ $L_{\text{\sun}}$.

In case of normal star-forming late-type galaxies, a combination of indicators that trace the dust-enshrouded and unobscured star formation are able to give a complete picture of the star formation activity. Since UV radiation is mainly originating in massive young stars, it is an ideal tracer for the unobscured star formation. Solely relying on this unobscured fraction will underestimate the true star formation activity, as starburst regions are often affected by strong attenuation from surrounding interstellar dust clouds. To acquire a complete picture of the rate at which stars are formed, the dust-enshrouded star formation must be traced as well. 
The total amount of enshrouded star formation can be traced by the total IR-luminosity.  But also several monochromatic SFR indicators have proven to be reliable estimators for the star formation activity. In particular, the mid-IR emission at 24$\mu$m shows a general correlation with the star formation rate \citep{2004A&A...428..409B, 2005ApJ...633..871C, 2005ApJ...632L..79W, 2006ApJ...650..835A, 2006ApJ...648..987P, 2007ApJ...667L.141R, 2007ApJ...666..870C, 2008A&A...479...83B, 2008ApJ...686..155Z, 2009ApJ...692..556R, 2009ApJ...700..161S, 2009ApJ...703.1672K}. 

In this paper, we investigate the correlation between [C{\sc{ii}}] luminosities and the SFR, estimated from the GALEX FUV and MIPS 24 $\mu$m or a combination of these SFR tracers. Based on the strength of these correlations, we will choose the SFR indicator showing the tightest correlation with [C{\sc{ii}}], which will serve as a reference SFR diagnostic in our analysis to calibrate the SFR-[C{\sc{ii}}] relation.
$\S$2 describes the sample selection and data acquisition. In $\S$3, we estimate the SFR either from the monochromatic 24$\mu$m luminosity or in combination with the GALEX FUV luminosity. Finally, $\S$4 discusses the reliability and applicability of [C{\sc{ii}}] as a star formation rate indicator and also addresses the nature of [C{\sc{ii}}] emission in galaxies. $\S$5 briefly summarizes our main conclusions.

\section{Sample and data}

\subsection{Sample selection}
The selection is based on the galaxy sample in \citet{2008ApJS..178..280B},  who assembled all galaxies with available [C{\sc{ii}}] data from the ISO archive. More specifically, we restrict our sample to those galaxies for which the [C{\sc{ii}}] line fluxes do not correspond to upper limits (i.e. non-detections) and which are classified in this paper as unresolved in the far-IR with respect to the $\sim$ 75$^{\prime \prime}$ \textit{ISO} LWS beam. This latter criterion implies that an aperture correction is not required for the [C{\sc{ii}}] line flux. Although a substantial fraction of the flux from an on-axis point source is diffracted out of the aperture beyond the diffraction limit at about 110 $\mu$m for all LWS detectors, these losses will be cancelled out in the calibration process, provided it is applied to point sources observed on-axis \citep{2003sws..book.....G}. 

From this sample of 153 unresolved galaxies, we retain all galaxies that have been observed in both GALEX FUV and MIPS 24$\mu$m bands, which gives us a final sample of 39 unresolved galaxies. 

Some galaxies have multiple [C{\sc{ii}}] data, corresponding to apertures taken at different positions. For every galaxy, we only include one data point, preventing the introduction of any bias in our sample. If the different apertures show some overlap in area, we choose the aperture corresponding to the [C{\sc{ii}}] flux with the smallest uncertainty. If the overlap is insignificantly small, we add all fluxes from the apertures taken at different positions. The fluxes from the different apertures for the GALEX FUV and MIPS\,24 $\mu$m are summed accordingly. This kind of flux summation has only been applied for two galaxies in our sample: NGC\,4651 and NGC\,7217.  

Since FUV data are affected by galactic extinction, we have to apply appropriate corrections.
At a galactic latitude $b$~=~11.$^{\circ}$2 \citep{1983MNRAS.204..811C}, NGC\,1569 suffers from a large amount of Galactic extinction. Inconsistencies among the different reported estimates for the V-band extinction (e.g. $A_{\text{V}}$ = 1.79 \citep{1988A&A...194...24I}; $A_{\text{V}}$ = 1.61 \citep{1997ApJ...482..765D}; $A_{\text{V}}$ = 2.32 \citep{1998ApJ...500..525S}), made us decide to remove NGC\,1569 from our sample, considering we want to avoid the introduction of any bias. This reduces our sample to 38 galaxies.

\subsection{GALEX data}
Far-ultraviolet (FUV, $\lambda$=1539\AA, $\Delta$$\lambda$=442\AA) observations of galaxies in our sample have been obtained from the GALEX GR4/5 public release. In order to derive
accurate FUV photometry we only used fields with integration times greater then 800 sec, obtained as part of the Nearby Galaxy Survey, the Medium Imaging Survey or Guest Investigator programs (70
galaxies from our unresolved sample of 153 galaxies have been observed by GALEX).
We performed aperture photometry on the intensity maps produced by the standard GALEX pipeline.
Details about the GALEX pipeline can be found in \citet{2007ApJS..173..682M}.
The GALEX FUV data have been corrected for galactic extinction according to \citet{1998ApJ...500..525S}, using the extinction relation obtained from \citet{1989ApJ...345..245C}. 

\subsection{MIPS 24$\mu$m data}
The 24 micron images were created from raw data produced by the
Multiband Imaging Photometer for Spitzer \citep[MIPS;][]{2004ApJS..154...25R} on
the Spitzer Space Telescope \citep{2004ApJS..154....1W}.  The raw data were
taken by a variety of programs using either the photometry mode (which
produces 5 times 5 arcmin maps) or the scan map mode (which produces
images that are typically 1 degree in length).  We used the MIPS Data
Analysis Tools \citep{2005PASP..117..503G} along with additional processing
steps to produce the final images. The individual data frames were
processed through droop correction (to remove excess signal in each
pixel proportional to the signal in the entire array), non-linearity
correction, dark current subtraction, scan mirror position-dependent
and position-independent flatfielding, and latent image removal steps.
Background emission from the zodiacal light and additional scattered
light that is related to the scan mirror position is then subtracted
out.  For objects observed in the scan map mode, the data from
separate astronomical observation requests were used to make mosaics
taken at individual epochs.  These images were then subtracted from
each other to identify asteroids, and the regions with asteroids were
then masked out.  Preliminary mosaics of each object were made using
all data for each object to identify pixels in individual frames with
values that are statistical outliers compared to other co-spatial
pixels in other frames.  These pixels were masked out, and then final
mosaics were made using an image scale of 1.5 arcsec pixel$^{-1}$.  Any
residual background in the images was measured in multiple regions
surrounding the optical disc and was then subtracted from the image.
The final images are then calibrated using the conversion factor given
by \citet{2007PASP..119..994E}, which has an uncertainty of 4$\%$.  The
FWHM of the PSF is 6 arcsec (Spitzer Science Center 2007).

\subsection{Flux determination}
We determine GALEX FUV and MIPS 24 $\mu$m fluxes within the same aperture as the $\sim$ 75$^{\prime \prime}$ \textit{ISO} LWS beam. Variation in the shape of the point spread function used by the different instruments should not be an issue, as the necessary corrections have been applied through the calibration on a point source.

The GALEX FUV and MIPS 24$\mu$m fluxes were determined using the source-extracting code SExtractor \citep{1996A&AS..117..393B}. To align the apertures with the exact same position at which the [C{\sc{ii}}] fluxes were determined, we have created a fake object in IRAF (using the task mkobjects) at the same position coordinates of the actual pointing of the [C{\sc{ii}}] observation. Feeding SExtractor with this artificially created map and the FUV or 24$\mu$m image, assures us that the FUV and 24$\mu$m fluxes are extracted within the same area on the sky as the [C{\sc{ii}}] fluxes.

\subsection{Spectral classification}
\label{sectie24}

\begin{figure}
  \centering
  \includegraphics[width=0.45\textwidth]{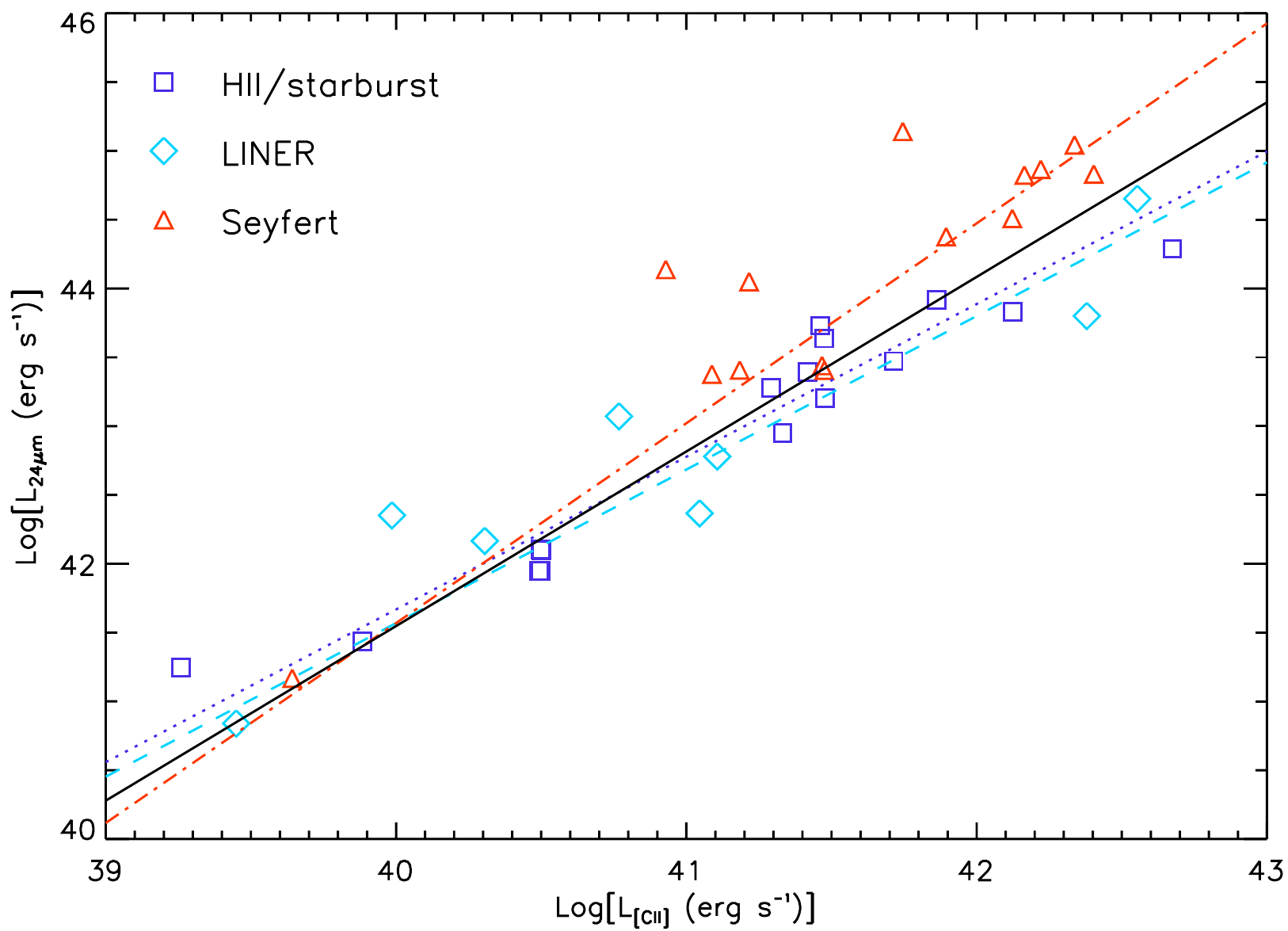}
  \includegraphics[width=0.45\textwidth]{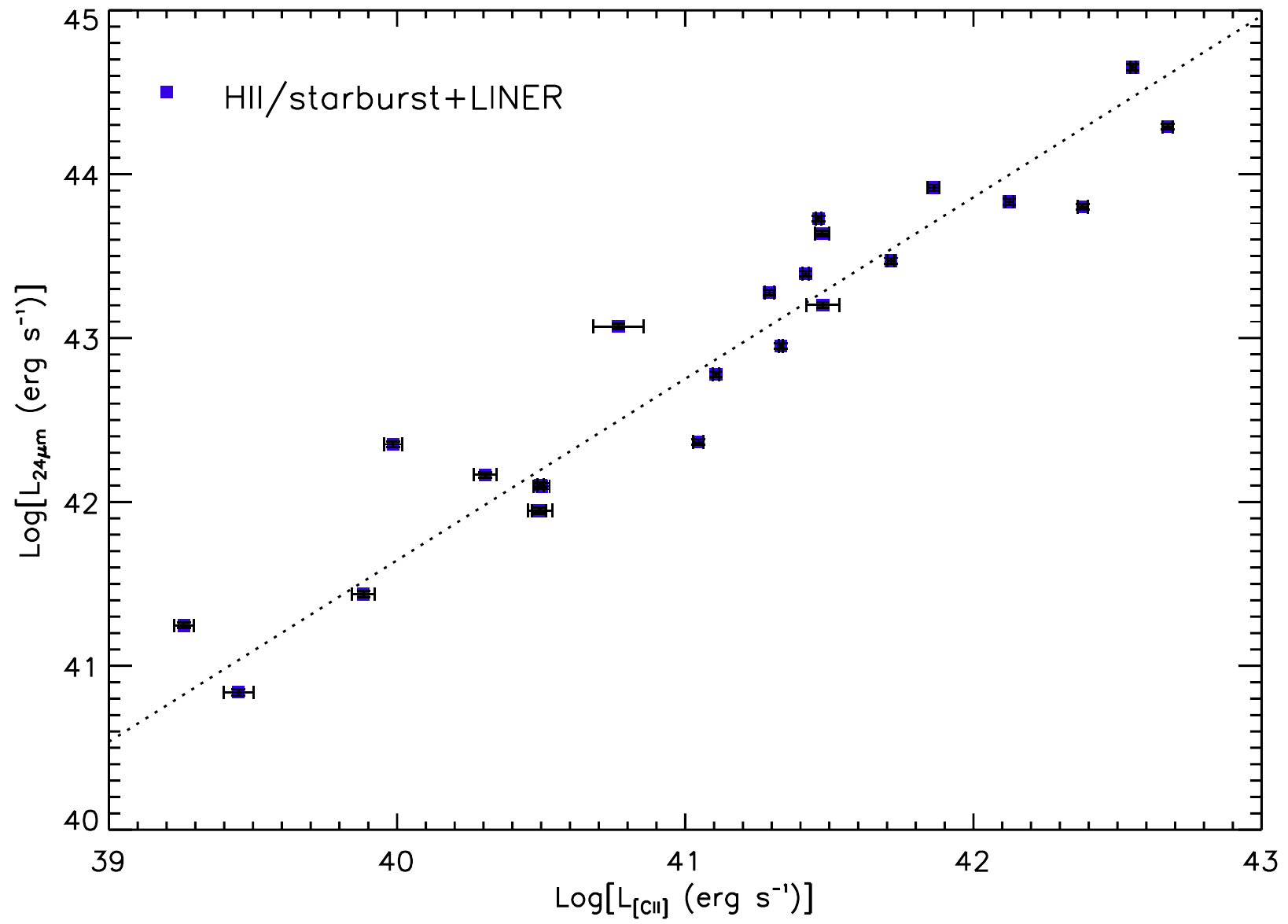}
  \caption{
$L_{[\text{C{\sc{ii}}}]}$ as a function of $L_{24\mu\text{m}}$. Upper panel: the three subsamples are represented  (blue squares = H{\sc{ii}}/starburst, cyan diamond = LINER, red triangle = Sy1/Sy2). For reasons of clarity, we have omitted the errorbars from this plot. The best fitting line through the data points are indicated with a blue dotted, cyan dashed and red dashed-dotted line for the H{\sc{ii}}/starburst, LINER and Seyfert subsample, respectively. The mean trend for the complete sample is represented by a black solid line. The data points have a dispersion of 0.51 dex around this mean trend. Lower panel: the reduced sample, combining H{\sc{ii}}, starburst and LINER galaxies. The dotted line represents the best fitting line through the data points with a dispersion of 0.31 dex.}
  \label{plot_lcii_24mu.gif}
\end{figure}

Considering we will calibrate the [C{\sc{ii}}] luminosity against the SFR, derived either from the monochromatic 24$\mu$m flux or in combination with the FUV flux, we have to make sure that the star-forming regions are the dominant contributor to this MIR emission. Therefore, we will classify our sample according to their nuclear spectral classification (see Table \ref{Table1}). We have made three different subdivisions. The first subgroup encompasses all objects resembling features typical of H{\sc{ii}} regions or starbursts. The second and third subset will include LINER and Seyfert galaxies, respectively.  
For this spectroscopic classification, we mainly rely on \citet{2010A&A...518A..10V}, who have composed a compilation of AGN host galaxies. For the identification of the remaining sources and the objects with an uncertain classification in \citet{2010A&A...518A..10V}, we have adopted the classification based on the optical spectra in \citet{1995ApJS...98..171V} and \citet{1997ApJS..112..315H}, who both use similar selection criteria in their classification procedure. Additionally, for some sources we used the classification from \citet{1986A&AS...66..335V}. Considering that the optical spectroscopic classification for FIR-luminous objects not always offers a clear distinction between starburst- or AGN-dominated nuclear activity, we additionally examine spectroscopic L-band data \citep{2004MSAIS...5..217R} and radio properties \citep{2006A&A...449..559B}, if available for those objects. Based on several L-band diagnostics, \citet{2004MSAIS...5..217R} found a dominant AGN-contribution to the energy output in the infrared for IRAS 19254-7245, IRAS 20551-4250 and one of the two optical nuclei in IRAS 23128-5919. \citet{2006A&A...449..559B} found from their radio observations that an AGN is the dominant power source in the nucleus of CGCG1510.8+0725. Four sources remain unclassified, but since they all host H{\sc{ii}} regions (Cartwheel: \citealt{2003ApJ...596L.171G}, NGC1317: \citealt{1996ApJS..105..353C}, NGC\,4189 and NGC\,4299: \citealt{1983AJ.....88..296H}), we assign them to the H{\sc{ii}}/starburst group.

Our sample of 38 galaxies contains 16 galaxies for which H{\sc{ii}} regions or starbursts dominate their central regions, while the remaining sources in our sample are host galaxies of an AGN (14 galaxies) or LINER (8 galaxies). For these different nuclear regimes, we will examine the power sources that mainly contribute to the 24 $\mu$m emission. Plotting $L_{[\text{C{\sc{ii}}}]}$ as a function of the MIPS\, 24$\mu$m luminosity for the three different subsets, we immediately deduce a significant contribution from the AGN in Seyfert galaxies to the 24$\mu$m emission (see the upper panels in Figure \ref{plot_lcii_24mu.gif}). The dispersion around the mean trend for the complete sample of galaxies is 0.51 dex in Figure \ref{plot_lcii_24mu.gif}. For galaxies classified as a LINER, starbursts appear to be the main power supply for the MIPS\, 24$\mu$m. Although the spectral line emission in some LINERs might also be generated by a quiescent AGN, this contribution does not appear to be significant for the LINER galaxies in our sample.  
Because of the insignificant AGN contribution to the MIR emission in LINERs, we perform our calibration analysis on a reduced sample that combines the two subsets (H{\sc{ii}} regions, starbursts and LINERs), which all have a dominant contribution from starburst to the MIR emission. This final sample consists of 24 galaxies of which 16 show features of H{\sc{ii}} regions and 8 other galaxies are classified as a LINER. Plotting again the same correlations for this final sample (see Figure \ref{plot_lcii_24mu.gif}, bottom panel), we can quantify that most of the scatter in previous plots was due to the contribution of an AGN to the 24$\mu$m emission. Indeed, the dispersion around the mean trend reduces to 0.31 dex in Figure \ref{plot_lcii_24mu.gif} (bottom panel).

\subsection{Sample description}
\begin{figure}
  \centering
  \includegraphics[width=0.45\textwidth]{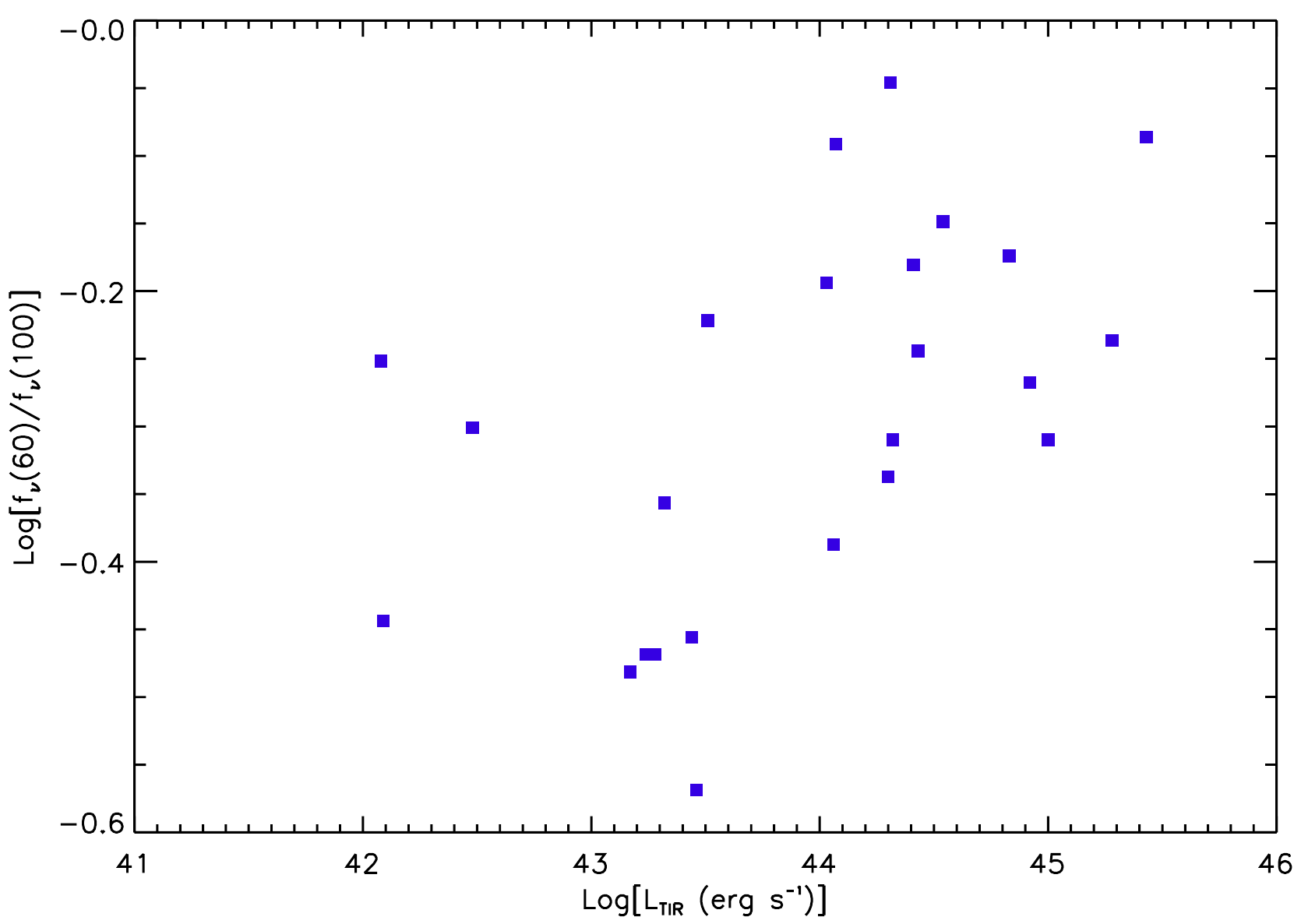}
  \caption{
The ratio $f_{\nu}$(60$\mu$m)-to-$f_{\nu}$(100$\mu$m), which is a proxy for the effective dust temperature, as a function of the total infrared luminosity $L_{\text{TIR}}$.}
  \label{plot_tir_ratio60100.pdf}
\end{figure}

The galaxies in our final sample span a range of almost 4 orders of magnitude in total infrared luminosity ($L_{\text{TIR}}$), from $\sim$ 1.2 $\times$ 10$^{42}$ to $\sim$ 2.7 $\times$ 10$^{45}$ erg s$^{-1}$ ($L_{\text{TIR}}$ was calculated based on equation (5) in \citealt{2002ApJ...576..159D}, relying on the  \textit{IRAS} 25, 60 and 100 $\mu$m fluxes). In this final sample of 24 galaxies (H{\sc{ii}} regions, starbursts and LINERs), five galaxies are classified as LIRG ($L_{\text{TIR}}$ $>$ 10$^{11}$ $L_{\text{\sun}}$). Figure \ref{plot_tir_ratio60100.pdf} demonstrates this wide range in $L_{\text{TIR}}$ and shows its dependence on the $f_{\nu}$(60$\mu$m)-to-$f_{\nu}$(100$\mu$m) ratio,  which is a proxy for the effective dust temperature. 
As most galaxies in our sample have high effective temperatures, the \textit{IRAS} 100$\mu$m flux samples the SED peak of dust emission rather well. This was also confirmed by \citet{2009ApJ...703.1672K}, as they found that the IRAS fluxes give a reliable representation of the total infrared luminosity for galaxies obeying $f_{\nu}$(60$\mu$m)/$f_{\nu}$(100$\mu$m) $>$ 0.4 . Only six galaxies in our sample have a $f_{\nu}$(60$\mu$m)-to-$f_{\nu}$(100$\mu$m) ratio below this value. 

Our sample covers a large range in distance, going from 3.9 to 139.1 Mpc. Four galaxies (NGC\,4189, NGC\,4293, NGC\,4299 and NGC\,7714) are in common with the sample that was used in \citet{2002A&A...385..454B}. 

Table \ref{Table1} gives an overview of all relevant information for our sample galaxies. We arranged their properties as follow:
\begin{itemize}
\item Column 1: galaxy name, NGC, IC, CGCG, UGC or IRAS, for the sample galaxies.
\item Column 2: distance [Mpc], derived from the Nearby Galaxies Catalogue \citep{1988Tully} and for other galaxies, from their recession velocity (from NED) assuming $H_{0}$ = 70 km s$^{-1}$ Mpc$^{-1}$.
\item Column 3: morphological type, adopted from \citet{1991rc3..book.....D}.  
\item Column 4, 5: classification of the nuclear spectrum: H{\sc{ii}} or starburst, LINER, Seyfert (1 or 2), and the corresponding references for this classification.
\item Column 6:  $f_{\nu}$(60$\mu$m)-to-$f_{\nu}$(100$\mu$m) ratio of the IRAS fluxes, a proxy for the effective dust temperature. 
\item Column 7: Total-IR luminosity [erg/s] for the whole galaxy, as calculated from equation (5) in \citet{2002ApJ...576..159D}.
\end{itemize}

Table 2 summarizes all relevant published data and measured fluxes within the $\sim$ 75$^{\prime \prime}$ \textit{ISO} LWS beam for the analysis in this paper. Summing up:
\begin{itemize}
\item Column 1: galaxy name, NGC, IC, CGCG, UGC or IRAS, for the sample galaxies.
\item Column 2, 3: Actual pointing position: right ascension (RA) and declination (DEC) from the [C{\sc{ii}}] observation, both in decimal degrees (from \citealt{2008ApJS..178..280B}). 
\item Column 4: [C{\sc{ii}}] line flux, in [fW/m$^{2}$], and the uncertainty on this flux measurement  (from \citealt{2008ApJS..178..280B}).
\item Column 5: GALEX FUV flux, [mJy], and the corresponding uncertainty. Both values have been corrected for galactic extinction, according to \citet{1998ApJ...500..525S} and \citet{1989ApJ...345..245C}.
\item Column 6: Spitzer MIPS 24 $\mu$m flux, [mJy], and the corresponding uncertainty, including both uncertainties in the flux extraction ($<$ 2$\%$) and the calibration (4$\%$, \citealt{2007PASP..119..994E}).
\end{itemize}

Although this final sample of 24 galaxies, gathering objects classified as H{\sc{ii}} region, starburst or LINER, is neither statistically significant nor representative for the whole of galaxies with diverging properties, this sample enables us to make a first preliminary analysis of the diagnostic capabilities of [C{\sc{ii}}] in tracing the star formation activity in galaxies. For a thorough analysis based on a more extensive sample of galaxies, we will have to wait until the completion of all [C{\sc{ii}}] surveys with the Herschel Space Observatory. Nevertheless, the broad range of optical and infrared luminosities covered in our sample enables us to infer the reliability of [C{\sc{ii}}] as a star formation indicator in star-forming galaxies spanning almost 4 orders of magnitude. 

\begin{savenotes}
\begin{table*}
\begin{minipage}{0.98\linewidth}
\begin{center}
\caption{Properties for the galaxies in our sample.}
\begin{tabular}{lcccccc}
  \hline\hline \\
  Name & Distance & Type\footnotemark[1] & Spectral type  & Ref\footnotemark[2]  &  IRAS\,60 / IRAS\,100 & log $L_{\text{TIR}}$\\ 
            & (Mpc) & & &  & & log (erg/s)\\
 (1) & (2) & (3) & (4) & (5) & (6) & (7) \\
 \\ \hline \\
 Cartwheel & 129.3 & RING  & H{\sc{ii}}/sb & no\footnotemark[3] & 0.46 & 44.30 \\
 NGC\,0520 & 27.8 & Pec & H{\sc{ii}}/sb & 3 & 0.66 & 44.41 \\
 NGC\,0625 & 3.9 & SB(s)m? edge-on & H{\sc{ii}}/sb & 1 & 0.56 & 42.08 \\
 NGC\,0660 & 11.8 & SB(s)a pec & LINER & 1, 2, 3 & 0.64 & 44.03\\
 NGC\,0695 & 139.1 & S0? pec  & H{\sc{ii}}/sb & 2  & 0.58 & 45.28 \\
 NGC\,0986 & 23.2 & SB(rs ab & H{\sc{ii}}/sb & 4 & 0.49 & 44.32 \\
 UGC\,02238 & 92.3 & Im? & LINER & 2 & 0.54 & 44.92 \\
 NGC\,1156 & 6.4 & IB(s)m & H{\sc{ii}}/sb & 3 & 0.50 & 42.48 \\
 NGC\,1266 & 31.3 & (R') SB(rs) 0$\wedge$0 pec: & LINER & 1, 2 & 0.81 & 44.07 \\
 NGC\,1275 & 75.2 & Pec & Seyfert & 1, 3 & 1.02 & 44.49 \\
 NGC\,1317 & 16.9 & SAB(r)a & H{\sc{ii}}/sb & no & 0.34 & 43.28 \\
 NGC\,1569 & 1.6 & IBm & H{\sc{ii}}/sb & 3 & 0.98 & 42.15 \\
 IRAS\,05189-2524 & 182.3 & & Seyfert & 1, 2 & 1.20 & 45.74 \\
 UGC\,03426 & 57.9 & S0: & Seyfert & 1 & 1.12 & 44.58 \\
 NGC\,2388 & 59.1 & S? & H{\sc{ii}}/sb & 2 & 0.67 & 44.83 \\
 NGC\,4041 & 22.7 & SA(rs)bc: & H{\sc{ii}}/sb & 3 & 0.41 & 44.06 \\
 NGC\,4189 & 16.8 & SAB(rs) cd? & H{\sc{ii}}/sb & no & 0.34 & 43.24 \\
 NGC\,4278 & 9.7 & E1-2 & LINER & 3 & 0.36 & 42.09 \\
 NGC\,4293 & 17.0 & (R) SB(s) 0/a & LINER & 3 & 0.44 & 43.32 \\
 NGC\,4299 & 16.8 & SAB(s)dm: & H{\sc{ii}}/sb & no & 0.33 & 43.17 \\
 NGC\,4490 & 7.8 & SB(s)d pec & H{\sc{ii}}/sb & 3 & 0.60 & 43.51 \\
 NGC\,4651 & 16.8 & SA(rs)c & LINER & 3 & 0.35 & 43.44 \\
 NGC\,4698 & 16.8 & SA(s)ab & Seyfert & 1, 3 & 0.31 & 42.65 \\
 IC\,4329A & 68.8 & SA0$\wedge$+: edge-on & Seyfert & 1 & 1.22 & 44.59 \\
 NGC\,5713 & 30.4 & SAB(rs)bc pec & H{\sc{ii}}/sb & 4 & 0.57 & 44.43 \\
 CGCG\,1510.8+0725 & 55.7 &  & Seyfert & 6 & 0.66 & 44.76 \\
 NGC\,6221 & 19.4 & SB(s)c & Seyfert & 1 & 0.50 & 44.37 \\
 NGC\,6240 & 104.8 & I0: pec & LINER & 1, 2 & 0.82 & 45.43 \\
 IRAS\,19254-7245 & 264.3 & & Seyfert & 1, 5 & 0.95 & 45.68 \\ 
 NGC\,6810 & 25.3 & SA(s)ab: & Seyfert & 4  & 0.52 & 44.28 \\
 IRAS\,20551-4250 & 184.1 & & Seyfert & 1, 5\footnotemark[4]  & 1.29 & 45.55 \\
 NGC\,7217 & 16.0 & (R)SA(r)ab & LINER & 1, 3 & 0.27 & 43.46 \\
 NGC\,7469 & 69.9 & (R')SAB(rs)a & Seyfert & 1, 2 & 0.74 & 45.25 \\
 IRAS\,23128-5919 & 191.0 & & Seyfert & 1, 5\footnotemark[5] & 0.98 & 45.58 \\
 NGC\,7552 & 19.5 & (R')SB(s)ab & H{\sc{ii}}/sb & 1 & 0.71 & 44.54 \\
 NGC\,7582 & 17.6 & (R')SB(s)ab & Seyfert & 1 & 0.67 & 44.25 \\
 NGC\,7714 & 36.9 & SB(s)b: pec & H{\sc{ii}}/sb & 1, 2 & 0.90 & 44.31 \\
 IRASF23365+3604 & 276.2 & & Seyfert & 1 & 0.84 & 45.74 \\
 NGC\,7771 & 61.1 & SB(s)b: pec & H{\sc{ii}}/sb & 2 & 0.49 & 45.00 \\
  \hline\hline 
\end{tabular}
\label{Table1}
\end{center}
\footnotemark[1]{ The galaxy type has been adopted from \citet{1991rc3..book.....D}. If no galaxy type is mentioned, there was no classification available in the literature.} \newline
\footnotemark[2]{ The references of the spectral type: 1: \citet{2010A&A...518A..10V}; 2: \citet{1995ApJS...98..171V}; 3: \citet{1997ApJS..112..315H}; 4: \citet{1986A&AS...66..335V}; 5: \citet{2004MSAIS...5..217R}  6: \citet{2006A&A...449..559B}} \newline
\footnotemark[3]{ "no'' indicates that no classification was found in the literature. These objects were classified in the H{\sc{ii}}/starburst group based on detected H{\sc{ii}} regions in their nuclei. Section \ref{sectie24} discusses the classification for these galaxies more into detail.} \newline
\footnotemark[4]{ Based on optical spectra IRAS\,20551-4250 was classified as a starburst galaxy \citep{2010A&A...518A..10V}, but according to L-band diagnostics a significant AGN contribution is present \citep{2004MSAIS...5..217R}.} \newline
\footnotemark[5]{ This galaxy has a pair of optical nuclei, which is probably a remnant of a merging process. \citet{2010A&A...518A..10V} identified this galaxy as H{\sc{ii}} region, while \citet{2004MSAIS...5..217R} reported that one nucleus shows AGN features and the other nucleus resembles a H{\sc{ii}} region. Since both nuclei fall within the ISO LWS beam, we have classified this object as a Seyfert galaxy, based on the AGN-contribution from at least one nucleus.} \newline

\end{minipage}
\end{table*}
\end{savenotes}

\begin{table*}
\centering
\caption{Sample galaxies: data}
\begin{tabular}{lccccccccr}
  \hline\hline \\
  Name & RA & DEC & [C{\sc{ii}}] & GALEX\,FUV & MIPS\,24 \\ 
 
            & (deg) & (deg) & (fW/m$^{2}$) & (mJy) & (mJy) \\
 (1) & (2) & (3) & (4) & (5) & (6) \\
 \\ \hline \\
   
 Cartwheel &  9.4171 & -33.7239 & 0.15 $\pm$ 0.02 & 1.980 $\pm$ 0.009 & 63.7 $\pm$ 2.7  \\
 NGC\,0520 &  21.1458	& +3.7950 & 2.83 $\pm$ 0.07 & 0.787  $\pm$ 0.009 & 2139.3 $\pm$ 85.7  \\
 NGC\,0625 & 23.7799 & -41.4357 & 1.00 $\pm$ 0.08 & 8.239 $\pm$ 0.016 & 776.0 $\pm$ 31.2 \\
 NGC\,0660 & 25.7595 & +13.6458 & 7.67 $\pm$ 0.19 & 0.386  $\pm$ 0.007 & 2889.9 $\pm$ 115.7  \\
 NGC\,0695 &27.8091 & +22.5825 & 2.04 $\pm$ 0.08 & 3.869  $\pm$ 0.003 & 672.4 $\pm$ 27.0 \\
 NGC\,0986 & 38.3922 & -39.0462 & 3.04 $\pm$ 0.11 & 0.964 $\pm$ 0.011 & 2355.9 $\pm$ 94.4  \\
 UGC\,02238 & 41.5726 & +13.0958 & 2.35 $\pm$ 0.09 & 0.448 $\pm$ 0.012 & 496.2 $\pm$ 20.0  \\
 NGC\,1156 & 44.9271 & +25.2375 & 1.56 $\pm$ 0.14 & 23.307  $\pm$ 0.115 & 447.1 $\pm$ 18.0  \\
 NGC\,1266 & 49.0032 & -2.4271 & 0.50 $\pm$ 0.10 & 0.216  $\pm$ 0.009 & 804.1 $\pm$ 32.3  \\
 NGC\,1275 & 49.9525 & +41.5105 & 1.16 $\pm$ 0.07 & 4.766  $\pm$ 0.014 & 2814.7 $\pm$ 112.7  \\
 NGC\,1317 & 50.6862 & -37.1027 & 0.91 $\pm$ 0.05 & 1.974  $\pm$ 0.013 & 207.2 $\pm$ 8.4 \\
 NGC\,1569 & 67.7046 & +64.8478 & 5.99 $\pm$ 0.16 & 105.217  $\pm$ 0.314 & 6495.8 $\pm$ 260.0 \\
 IRAS\,05189-2524 & 80.2558 & -25.3624 & 0.14 $\pm$ 0.02 & 0.126  $\pm$ 0.004 & 2786.6 $\pm$ 111.6 \\
 UGC\,03426 & 93.9016	& +71.0376 & 0.41 $\pm$ 0.02 & 0.662 $\pm$ 0.020 & 2238.1 $\pm$ 89.7 \\
 NGC\,2388 & 112.2229	& +33.8182 &1.74 $\pm$ 0.08 & 0.135 $\pm$ 0.005 & 1585.2 $\pm$ 63.5 \\
 NGC\,4041 & 180.5501	& +62.1363 & 3.48 $\pm$ 0.05 & 1.880 $\pm$ 0.001 & 1159.7 $\pm$ 46.5 \\
 NGC\,4189 & 183.4465	& +13.4240 & 0.94 $\pm$ 0.06 & 2.243 $\pm$ 0.001 & 295.7 $\pm$ 12.0 \\ 
 NGC\,4278 & 185.0276	& +29.2824 & 0.25 $\pm$ 0.03 & 0.883 $\pm$ 0.010 & 49.1 $\pm$ 2.1  \\ 
 NGC\,4293 & 185.3050	& +18.3844 & 0.28 $\pm$ 0.02 & 0.111 $\pm$ 0.003 & 520.1 $\pm$ 20.9  \\
 NGC\,4299 & 185.4189	& +11.5012 & 0.93 $\pm$ 0.09 & 8.996 $\pm$ 0.030 &  210.4 $\pm$ 8.5   \\
 NGC\,4490 & 187.6536	& +41.6398 & 4.32 $\pm$ 0.11 & 10.248 $\pm$ 0.022 & 1394.8 $\pm$ 55.9  \\
 NGC\,4651 & 190.9271 & +16.3942 & 2.03 $\pm$ 0.08 & 4.048 $\pm$ 0.015 &  383.1 $\pm$ 15.5 \\
NGC\,4651 & 190.9442 &  +16.3972 & 0.50 $\pm$ 0.06 & 1.533 $\pm$ 0.010 &  79.6 $\pm$ 3.3 \\
NGC\,4651 & 190.91 & +16.3914 & 0.76 $\pm$ 0.09 & 1.732 $\pm$ 0.010 &  88.0 $\pm$ 3.6 \\
NGC\,4651 & total from all positions  & & 3.29 $\pm$ 0.13 & 7.314  $\pm$ 0.021 & 550.7 $\pm$ 22.2 \\
 NGC\,4698 & 192.0969	& +8.4875 & 0.13 $\pm$ 0.02 & 0.289 $\pm$ 0.004 & 34.9 $\pm$ 1.5 \\
 IC\,4329A & 207.3304 &	-30.3095 & 0.15 $\pm$ 0.02 & 0.073 $\pm$ 0.003 & 1945.3 $\pm$ 77.9 \\
 NGC\,5713 & 220.0478	& -0.2905 & 4.68 $\pm$ 0.13 & 4.061 $\pm$ 0.020 & 2144.6 $\pm$ 85.9  \\
 CGCG\,1510.8+0725 & 228.3053	& +7.2265 & 0.33 $\pm$ 0.05 & 0.086 $\pm$ 0.004 & 514.3 $\pm$ 20.7  \\
 NGC\,6221 & 253.1913	& -59.2167 & 6.64 $\pm$ 0.25 & 4.527 $\pm$ 0.063 & 4528.1 $\pm$ 181.3 \\
 NGC\,6240 & 253.2450	& +2.4013 & 2.72 $\pm$ 0.06 & 0.637 $\pm$ 0.004 & 2735.4 $\pm$ 109.5 \\
 IRAS\,19254-7245 & 292.8400 & -72.6319 & 0.26 $\pm$ 0.07 & 0.162 $\pm$ 0.005 & 1059.0 $\pm$ 42.5 \\ 
 NGC\,6810 & 295.8915	& -58.6556 & 3.83 $\pm$ 0.13 & 0.553 $\pm$ 0.010 & 2882.1 $\pm$ 115.4 \\
 IRAS\,20551-4250 & 314.6116 & -42.6518 & 0.41 $\pm$ 0.03 & 0.600 $\pm$ 0.006 & 1450.9 $\pm$ 58.2 \\
 NGC\,7217 & 331.9775	& +31.35833 & 0.66 $\pm$ 0.06 & 1.557 $\pm$ 0.008 & 197.9 $\pm$ 8.0 \\
 NGC\,7217 & 331.95833 & +31.35972 & 0.62 $\pm$ 0.07 & 1.331 $\pm$ 0.008 & 185.2 $\pm$ 7.5 \\
 NGC\,7217 & total from all positions& & 1.28 $\pm$ 0.09 & 2.888 $\pm$ 0.012 & 383.1 $\pm$ 15.5  \\
 NGC\,7469 & 345.81417 & +8.87361 & 2.27 $\pm$ 0.03 & 6.963 $\pm$ 0.020 & 4421.9 $\pm$ 177.0 \\
 IRAS\,23128-5919 & 348.9454 & -59.0544 & 0.58 $\pm$ 0.05 & 0.557 $\pm$ 0.005 & 1242.2 $\pm$ 49.8  \\
 NGC\,7552 & 349.045 & -42.5844 & 6.37 $\pm$ 0.15 & 3.746 $\pm$ 0.011 &  9443.8 $\pm$ 377.9  \\
 NGC\,7582 & 349.5988 & -42.3703  & 4.12 $\pm$ 0.13 & 0.900 $\pm$ 0.006 & 5516.4 $\pm$ 220.8  \\
 NGC\,7714 & 354.0612	& +2.1550 & 1.83 $\pm$ 0.10 & 7.893 $\pm$ 0.018 & 2131.7 $\pm$ 85.4 \\
 IRASF23365+3604 & 354.7554 & +36.3528 & 0.16 $\pm$ 0.02 & 0.197 $\pm$ 0.006 & 585.2 $\pm$ 23.5  \\
 NGC\,7771 & 357.8534	& +20.1119 & 2.98 $\pm$ 0.09 & 0.906 $\pm$ 0.012 & 1216.5 $\pm$ 48.8  \\
  \hline\hline
\end{tabular}
\label{Table2}
\end{table*}

\section{Reliable reference SFR tracer}
In order to compare the different star formation relations, it is important that they are calibrated in the same way. Therefore we will relate all relations to the same initial mass function (IMF). This reference IMF will be the \citet{2001MNRAS.322..231K} IMF, characterized by a power law $\xi$(m) $\propto$ m$^{-\alpha}$ with slope $\alpha$ = 2.3 for 0.5-100 M$_{\text{\sun}}$ and $\alpha$ = 1.3 for 0.1-0.5 M$_{\text{\sun}}$. We prefer the \citet{2001MNRAS.322..231K} IMF rather than the more commonly used \citet{1955ApJ...121..161S} IMF (with a single slope $\alpha$ = 2.35 for 0.1-100 M$_{\text{\sun}}$), since the former better correlates with IMF observations in the Galactic field \citep{2003PASP..115..763C,2003ApJ...598.1076K}.

Analyzing the correlations between $L_{[\text{C{\sc{ii}}}]}$ and the SFR requires a complete threefold approach:
\begin{itemize}
\item We first analyze the simple relation of the SFR as a function of the [C{\sc{ii}}] luminosity. Although this approach is most straightforward, the data might be biased towards larger distances. Since the distances to the galaxies in our sample range from 3.9 to 139.1 Mpc, we have to eliminate the distance bias in this luminosity versus luminosity relation.
\item Plotting SFR/4$\pi D^{2}$ as a function of the [C{\sc{ii}}] flux, the distance bias is removed. If a true [C{\sc{ii}}]-SFR correlation is present, it will be confirmed in this plot.   
\item The dispersion around the mean trend is best quantified in a plot of the [C{\sc{ii}}] luminosity versus the residual of the best fitting relation between $L_{[\text{C{\sc{ii}}}]}$ and the SFR. The SFR relation showing the tightest correlation with the [C{\sc{ii}}] luminosity will be characterized by the lowest dispersion in the residual plots. 
\end{itemize}

\subsection{SFR calibrated against TIR}
\label{sectiontir}
Taking into account that the 24$\mu$m emission from a galaxy better correlates with the SFR as opposed to monochromatic FIR measurements at longer wavelengths (70 and 160$\mu$m) \citep{2010ApJ...714.1256C}, we have only estimated the SFR, which is used for the final calibration of the SFR-[C{\sc{ii}}] relation, from 24$\mu$m data (or FUV luminosities which have been corrected for attenuation based on 24$\mu$m data) in this work. To obtain a second independent SFR diagnostic, we have investigated whether TIR luminosities for these galaxies could be used as a calibrator. Since we only have IRAS fluxes covering the whole galaxy for all sample objects to estimate the TIR luminosity, we checked whether an aperture correction was required. Although the galaxies in our sample were selected based on the criterion that they are unresolved in the FIR with respect to the \textit{ISO} LWS beam, many of them are resolved within the \textit{ISO} beam in optical data. Therefore, we visually inspected the MIPS 24$\mu$m images and could indeed quantify that also in the MIR a significant fraction of the total flux for several galaxies fell outside the \textit{ISO} beam aperture. Subsequently, we verified the availability of MIPS 70 and 160$\mu$m data in the Spitzer archive, to determine the correct TIR luminosities which correspond to the portion of the flux that falls within the same \textit{ISO} beam. Considering that only 28 of the 38 sample galaxies have full coverage in all Spitzer bands, and some of those galaxies were only observed in small fields which does not allow proper convolution with the appropriate kernel to match the PSF's in all Spitzer bands, we decided not to use these Spitzer data. Moreover, the varying sizes of the sample galaxies would imply a different aperture correction (some are point sources and others are clearly more extended), introducing an additional bias to the TIR luminosity derived from these Spitzer data.
While the MIPS 24$\mu$m clearly demonstrate that the majority of our sample galaxies requires some kind of aperture correction, we have verified that the data currently at hand are not able to properly correct for these aperture issues. Any empirical correction can significantly bias the correlations between the [C{\sc{ii}}] luminosity and the SFR derived from this aperture-corrected TIR luminosity and does not allow us to draw any firm conclusion. Therefore, this paper only investigates the correlation between [C{\sc{ii}}] and the SFR, when estimated from the 24$\mu$m or attenuation-corrected FUV flux.

\subsection{SFR calibrated against MIPS 24 $\mu$m}
In this section, we predict the star formation rate solely relying on the MIR emission at 24$\mu$m, to trace the dust-enshrouded fraction of star formation.

First, we have to choose an appropriate star formation relation.
Several SFR relations based on 24 $\mu$m emission have been published. Some authors claim a linear relation between the SFR and $L_{24\mu\text{m}}$ \citep{2005ApJ...632L..79W, 2008ApJ...686..155Z}, while others argue that only a non-linear relation can account for the increasing 24$\mu$m emission in more IR-luminous galaxies which form stars at a higher rate \citep{2005ApJ...632L..79W, 2006ApJ...648..987P, 2006ApJ...650..835A, 2007ApJ...666..870C, 2007ApJ...667L.141R, 2008ApJ...686..155Z, 2009ApJ...692..556R, 2010ApJ...713..626B}. 
As some galaxies in our sample are situated at the high luminosity end, we should be aware that this non-linear effect might affect our results. 
We choose to estimate the SFR from the relation in \citet{2009ApJ...692..556R}. 

\citet{2009ApJ...692..556R} use a combination of a linear trend at lower TIR-luminosities and a non-linear relation above a threshold of $L_{\text{TIR}}$ $>$ 10$^{11}$ $L_{\odot}$ or $L_{24\mu\text{m}}$ $>$ 1.3 $\times$ 10$^{10}$ $L_{\odot}$.
Both relations were derived from a sample of dusty, luminous star-forming galaxies and hold up to $L_{\text{TIR}}$ = 2 $\times$ 10$^{12}$  $L_{\odot}$, which is higher than the most luminous galaxy in our sample (NGC\,6240; 7.0 $\times$ 10$^{11}$ $L_{\odot}$).
The overlap between our range in TIR luminosities and the similar characteristics of both samples make us confident that this relation gives a good indication of the true star formation rate in our sample galaxies.
The two relations from \citet{2009ApJ...692..556R} are (assuming our reference \citealt{2001MNRAS.322..231K} IMF):
\begin{equation}
\text{SFR} = 2.04 \times 10^{-43} L_{24\mu m} 
\end{equation}
for 4 $\times$ 10$^{42}$ $\la$ $L_{24\mu\text{m}}$ $\la$ 5 $\times$ 10$^{43}$ erg s$^{-1}$ and
\begin{equation}
\text{SFR} = 2.04 \times 10^{-43} \times L_{24\mu m} (2.03 \times 10^{-44}  L_{24\mu m})^{0.048}
\end{equation} 
for $L_{24\mu\text{m}}$ $>$ 5 $\times$ 10$^{43}$ erg s$^{-1}$, where SFR and  $L_{24\mu\text{m}}$ have units M$_{\text{\sun}}$ yr$^{-1}$ and erg s$^{-1}$, respectively.

\begin{figure}
  \centering
  \includegraphics[width=0.45\textwidth]{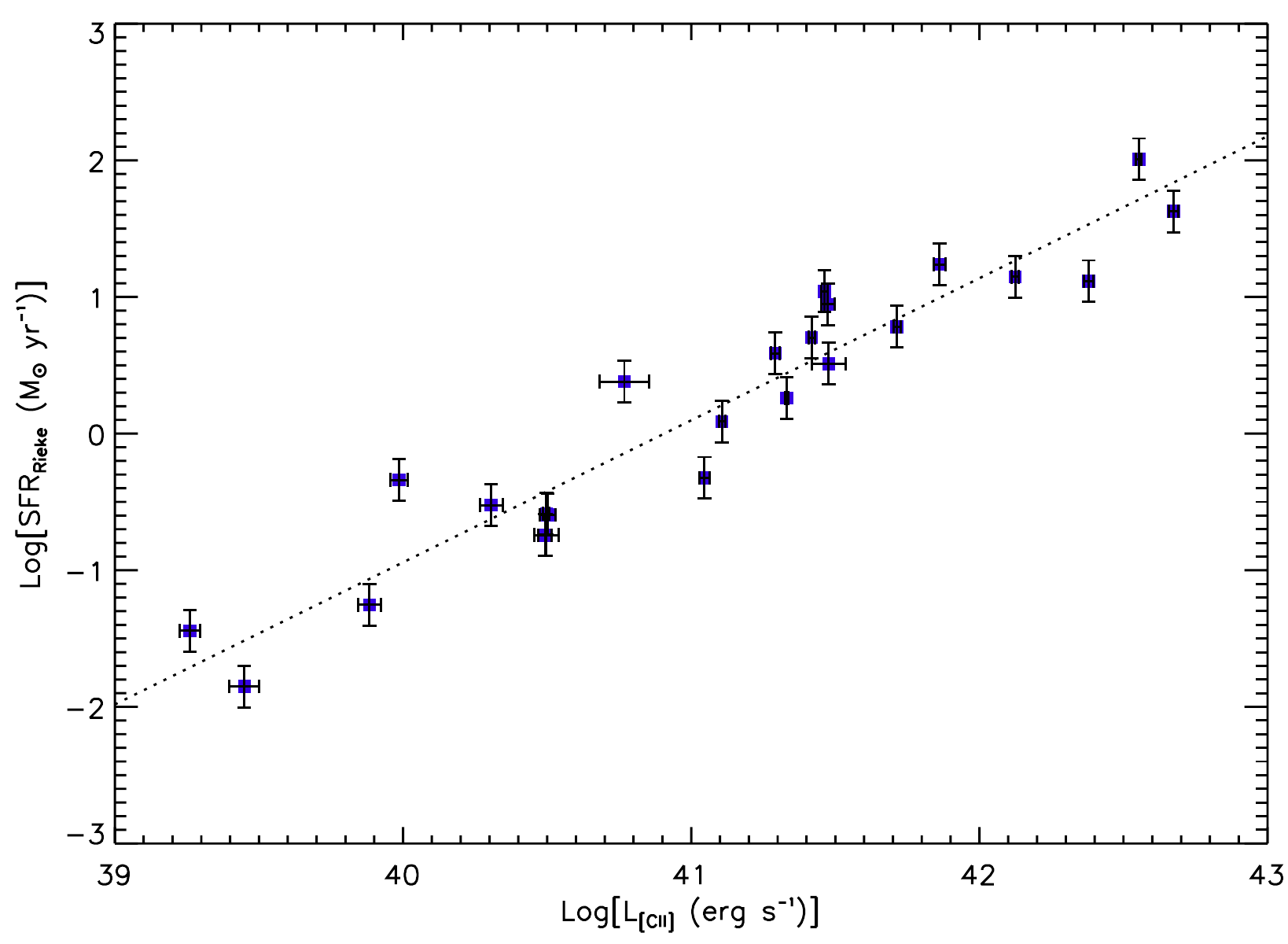} \newline
  \includegraphics[width=0.45\textwidth]{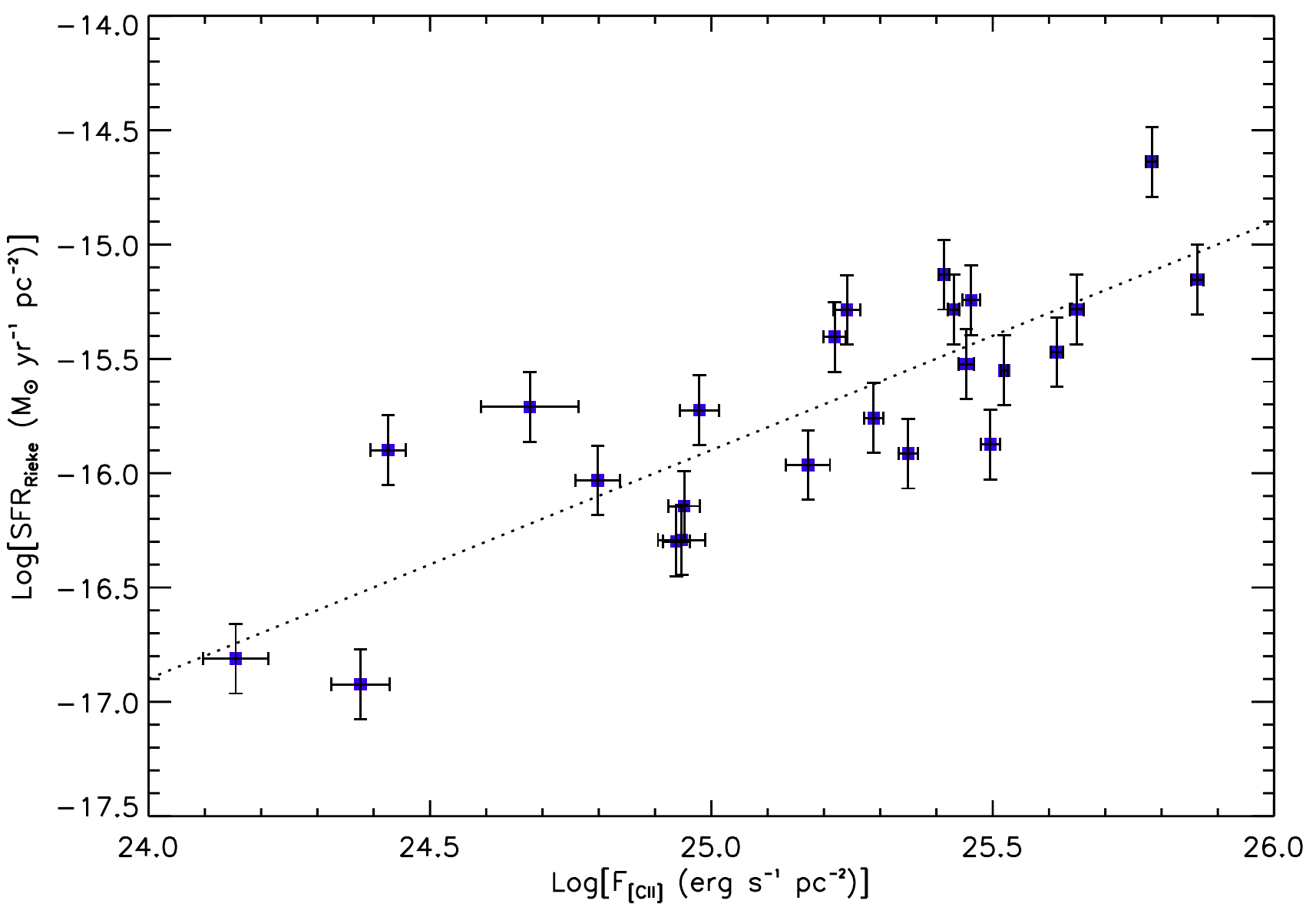} \newline
  \includegraphics[width=0.45\textwidth]{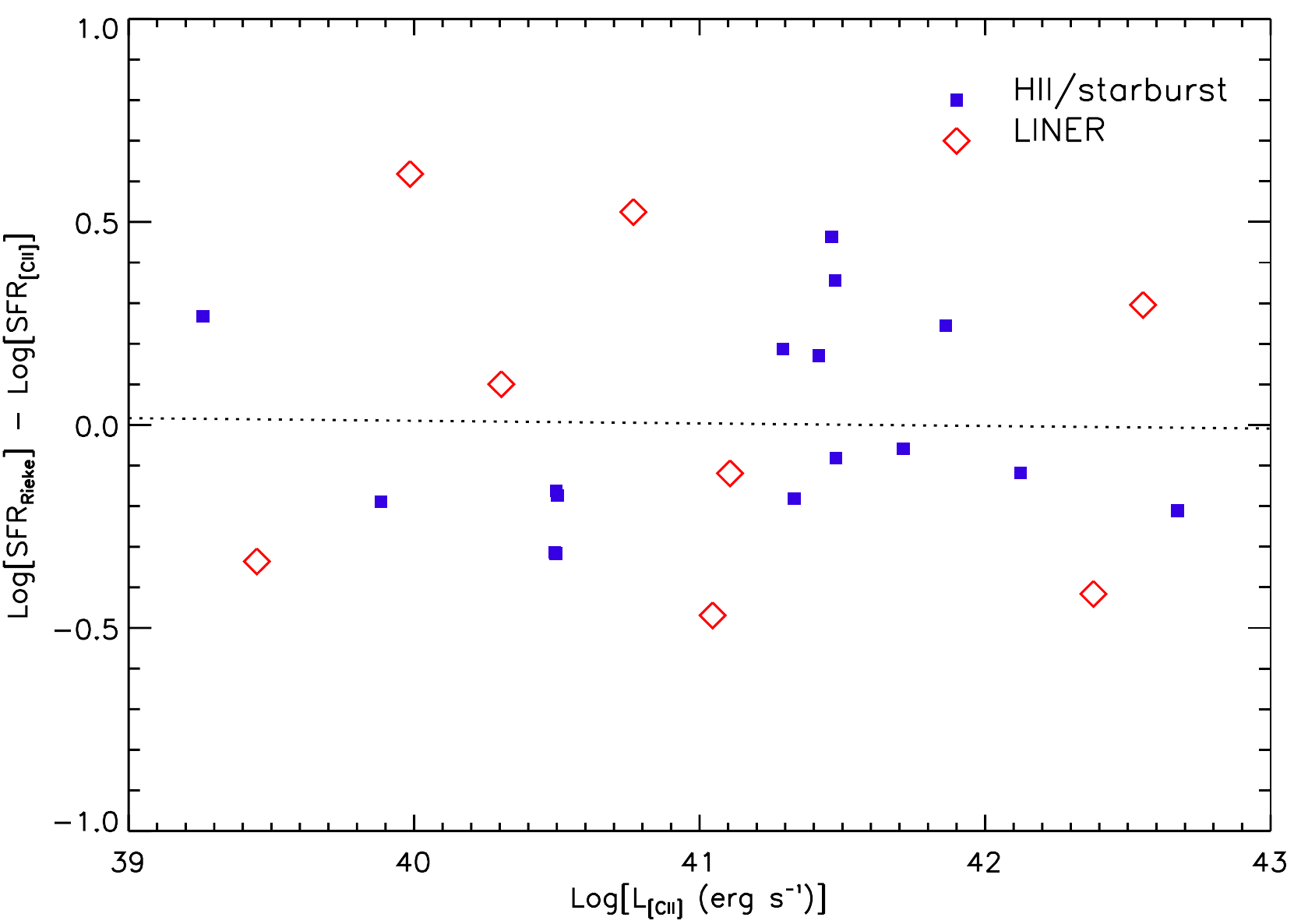} 
  \caption{
Upper panel: SFR$_{24\mu m}$ (SFR relation from \citet{2009ApJ...692..556R}) as a function of $L_{[\text{C{\sc{ii}}}]}$, central panel: SFR$_{24\mu m}/4\pi D^2$  as a function of [C{\sc{ii}}] flux, bottom panel: residuals of the SFR$_{24\mu m}$-$L_{[\text{C{\sc{ii}}}]}$ plot. In this residual plot, H{\sc{ii}}/starburst and LINER galaxies are indicated as blue filled squares and red diamonds, respectively.}
  \label{plot_sfrrieke_lcii.gif}
\end{figure}

After estimating the SFR from these relations, we verify the strength of the correlation between $L_{[\text{C{\sc{ii}}}]}$ and the SFR through a three-pronged approach.
First, we plot the derived SFR as a function of $L_{[\text{C{\sc{ii}}}]}$ (see Figure \ref{plot_sfrrieke_lcii.gif}, upper panel).
The best fitting line through the data points corresponds to:
\begin{equation}
\label{bestrieke}
\log \text{SFR}_{24\mu m} =  1.040 \log L_{\text{[C{\sc{ii}}]}} - 42.535
\end{equation}
where the units of SFR$_{24\mu\text{m}}$ and $L_{[\text{C{\sc{ii}}}]}$ are in M$_{\text{\sun}}$ yr$^{-1}$ and erg s$^{-1}$, respectively. 
The 1$\sigma$ dispersion of the individual galaxies around the best fitting line is 0.30 dex. 
This 1$\sigma$ dispersion has been calculated as the standard deviation of the logarithmic distance from the data points to the best fitting line weighted by the error on the individual quantities. Since the observed trend in this plot of luminosity versus SFR (and thus indirectly luminosity) might be biased by the range in distances among our sample galaxies, we remove this distance bias in the central panel of Figure \ref{plot_sfrrieke_lcii.gif}. In this figure, we plot the SFR-to-4$\pi$D$^2$ ratio as a function of the [C{\sc{ii}}] flux. Since this distance-independent plot still shows a similar correlation between [C{\sc{ii}}] and the SFR and the dispersion 0.30 dex is of the same order as in the first panel, we safely conclude that the observed correlation in the upper panel of Figure \ref{plot_sfrrieke_lcii.gif} is not due to a distance bias in the data set. 
The dispersion around the mean trend (see Equation \ref{bestrieke}) is better represented in the bottom panel of Figure \ref{plot_sfrrieke_lcii.gif}, which shows the residual plot. 

\subsection{SFR calibrated against GALEX FUV + MIPS 24 $\mu$m}
Before the GALEX FUV flux can be used as a direct quantifier of the star formation activity in a galaxy, a correction for internal dust attenuation has to be applied. The specific amount of extinction that is affecting the UV light is difficult to quantify, because different galaxies will be characterized by different extinction laws and a variation of stellar and dust distributions. 
In this section, we will rely on SFR relations that use a combination of the GALEX FUV flux and the monochromatic MIR emission at 24$\mu$m to obtain a complete picture of the star formation activity.

Two SFR relations have been reported, making use of both photometric data.
First, \citet{2008ApJ...686..155Z} derived a linear relation to estimate the attenuation-corrected FUV-flux based on the observed FUV and 24$\mu$m data:
\begin{equation}
\label{zhu}
L_{\text{FUV,corr}} [L_{\sun}]  = L_{\text{FUV,obs}}[L_{\sun}]  + 6.31 L_{24\mu \text{m}} [L_{\sun}]. 
\end{equation}
For the derivation of this relation, \citet{2008ApJ...686..155Z} relied on a sample of 187 star-forming (non-AGN) galaxies. 

Another relation, reported in \citet{2008AJ....136.2782L}, also computes the SFR from both FUV and 24$\mu$m data.
Since a large fraction of our sample galaxies are characterized by a much higher star formation activity with respect to the sample in \citet{2008AJ....136.2782L}, we will apply the attenuation correction as reported in \citet{2008ApJ...686..155Z} and rely on the SFR relation in \citet{2009ApJ...703.1672K}:
\begin{equation}
\label{kenn}
\text{SFR} = 0.88 \times 10^{-28} L_{\text{FUV,corr}},
\end{equation} 
where the units of $L_{\text{FUV,corr}}$ are in erg s$^{-1}$ Hz$^{-1}$, to derive a star formation rate from the extinction-corrected FUV data.
This relation was derived for a sample of normal galaxies in the present-day universe and most star-forming galaxies
out to redshifts z $\sim$ 1. With an upper limit of $\log$ $L_{\text{TIR}}$[$L_{\text{\sun}}$] $\sim$ 11.9, the relation is not immediately representive for the most dust-obscured LIRGs or ULIRGs found in the present-day universe (but this is not applicable for galaxies in our non AGN-dominated sample).
 
\begin{figure}
  \centering
  \includegraphics[width=0.45\textwidth]{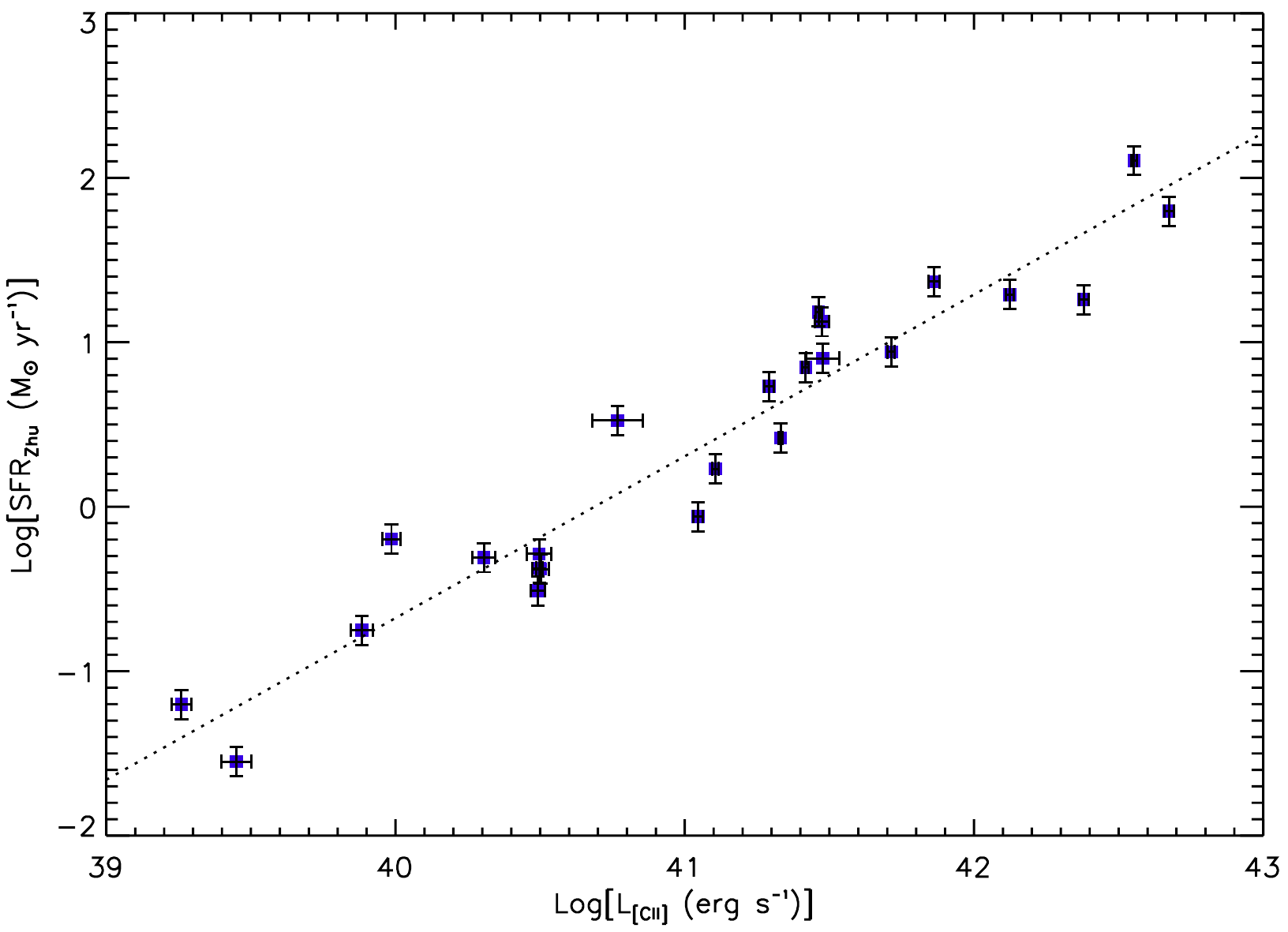} \newline
  \includegraphics[width=0.45\textwidth]{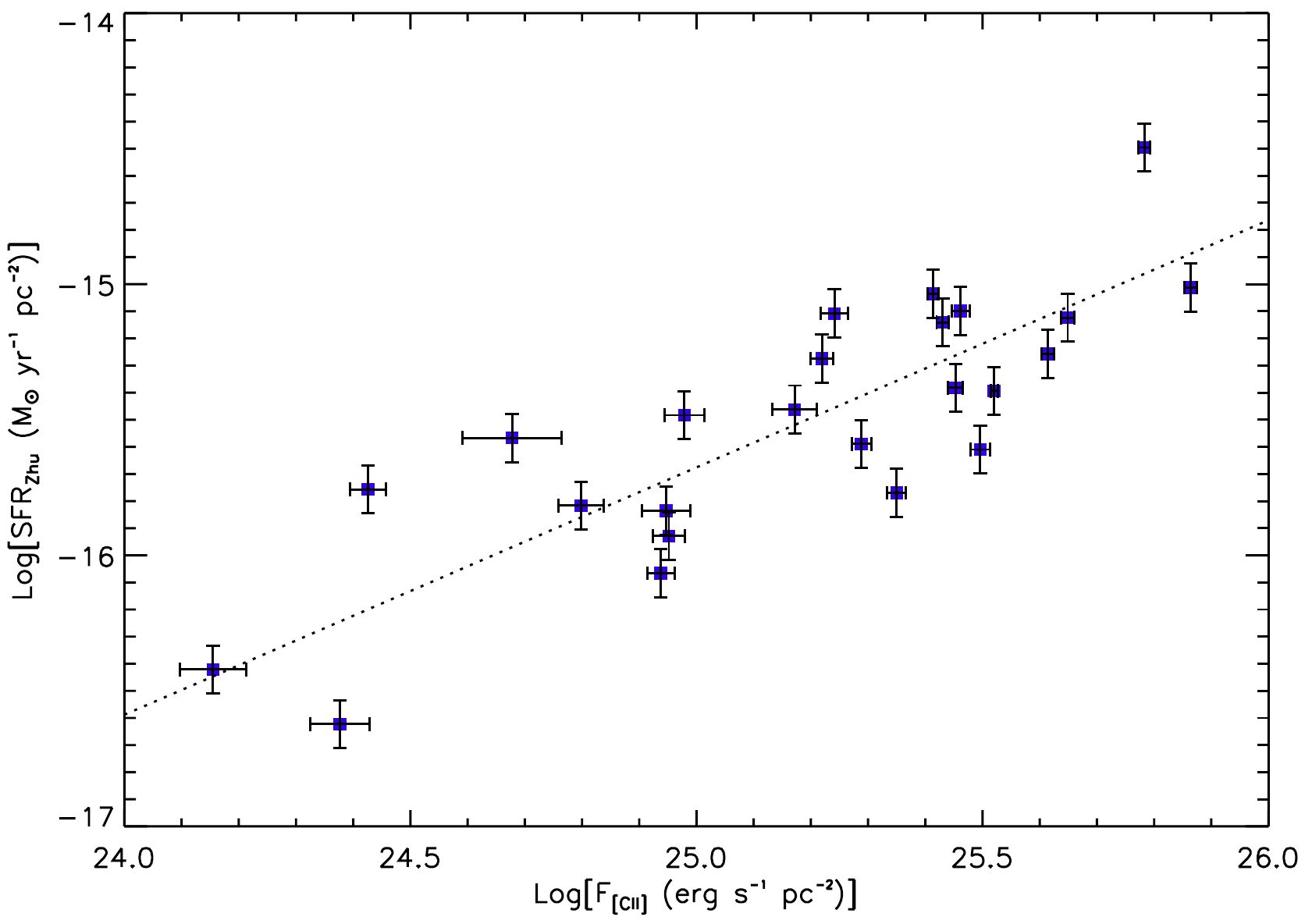} \newline
  \includegraphics[width=0.45\textwidth]{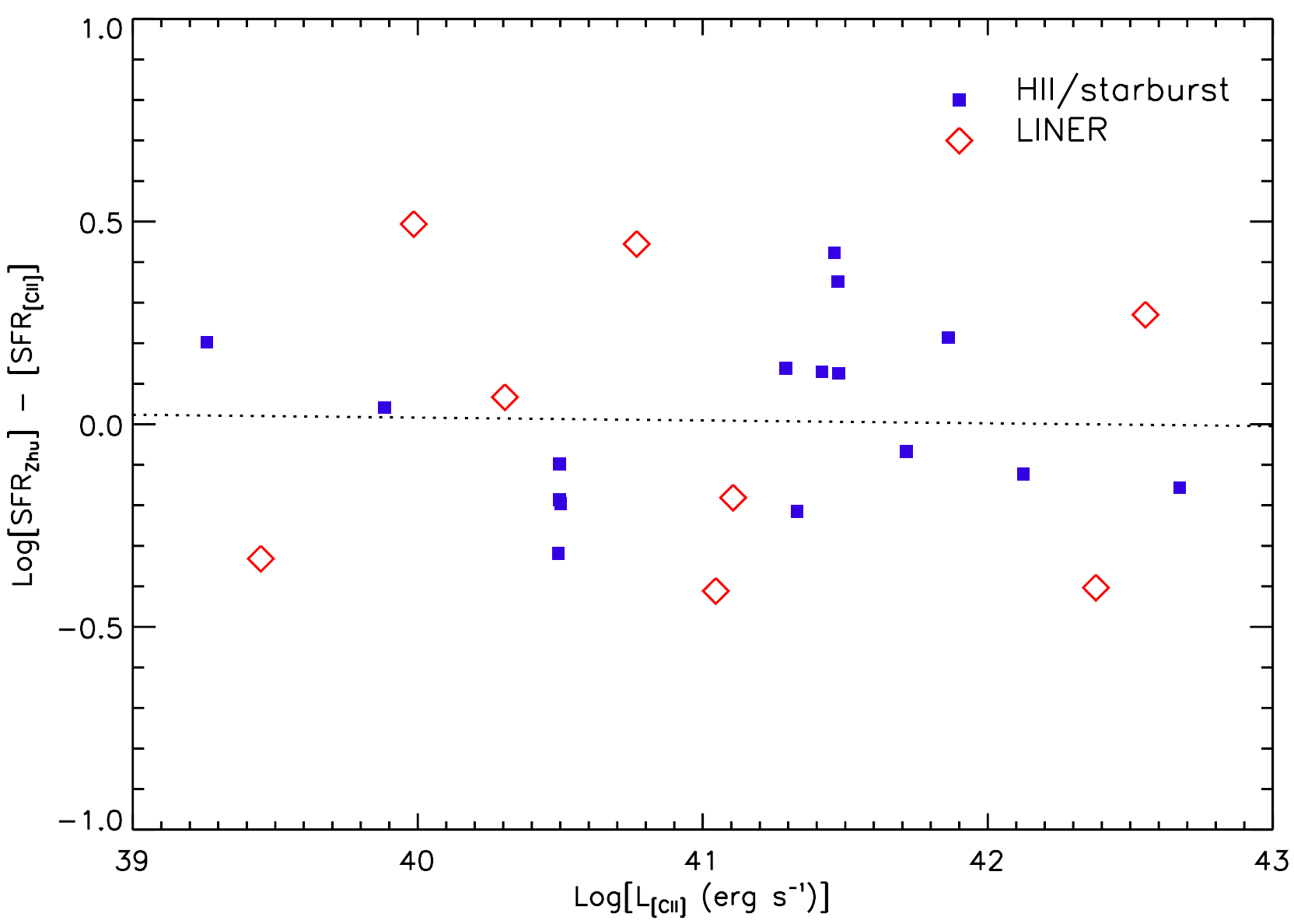} 
  \caption{
SFR$_{FUV+24\mu m}$ (attenuation correction and SFR relation from \citet{2008ApJ...686..155Z} and \citet{2009ApJ...703.1672K}, respectively) as a function of $L_{[\text{C{\sc{ii}}}]}$, central panel: SFR$_{FUV+24\mu m}/4\pi D^2$  as a function of [C{\sc{ii}}] flux, bottom panel: residuals of the SFR$_{FUV+24\mu m}$-$L_{[\text{C{\sc{ii}}}]}$ plot. In this residual plot, H{\sc{ii}}/starburst and LINER galaxies are indicated as blue filled squares and red diamonds, respectively.}
  \label{plot_sfrzhu_lcii.gif}
\end{figure}

Figure \ref{plot_sfrzhu_lcii.gif} shows the correlation between $L_{[\text{C{\sc{ii}}}]}$ and the SFR$_{\text{FUV+24}\mu \text{m}}$ derived in this way.
The best fitting line through all data points is given by:
\begin{equation}
\label{FUV+24}
\log \text{SFR}_{\text{FUV+24}\mu\text{m}} = 0.983 \log L_{\text{[C{\sc{ii}}]}} - 40.012
\end{equation}
where the units of SFR$_{\text{FUV}+24\mu\text{m}}$ and $L_{[\text{C{\sc{ii}}}]}$ are in M$_{\text{\sun}}$ yr$^{-1}$ and erg s$^{-1}$, respectively. The spread around the mean trend is represented in the bottom panel of Figure \ref{plot_sfrzhu_lcii.gif} and can be quantified by a 1$\sigma$ dispersion of 0.27 dex. The central 
panel of Figure \ref{plot_sfrzhu_lcii.gif} again confirms that the observed trend is not due to a distance bias in our sample, since the 1$\sigma$ spread around the mean trend (0.26 dex) does not differ in this distance independent relation.

\begin{figure}
  \centering
  \includegraphics[width=0.45\textwidth]{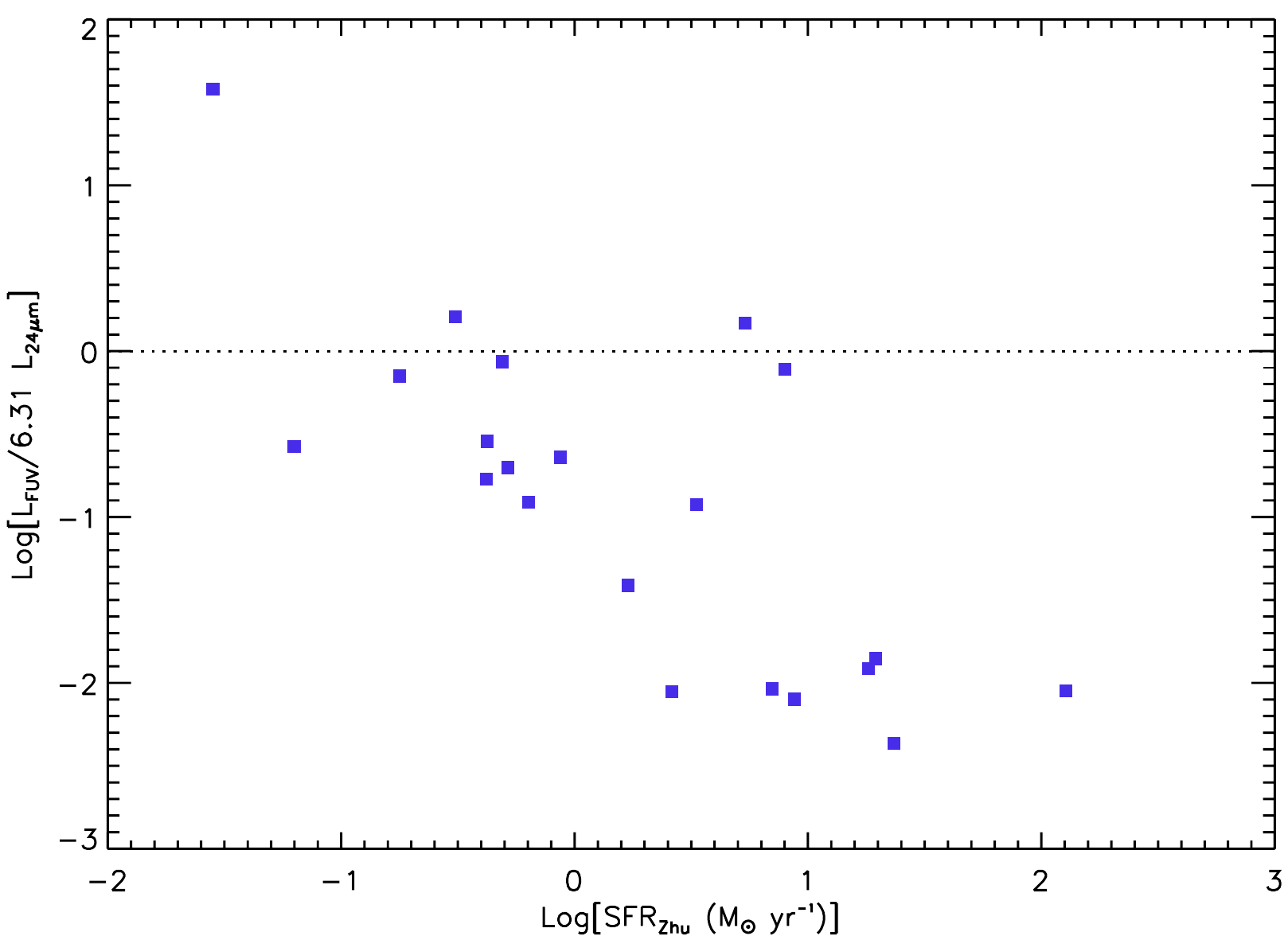}
  \caption{
The ratio $L_{\text{FUV}}$ / 6.31 $L_{24\mu\text{m}}$ as a function of the total SFR, as derived from formula \ref{zhu} and \ref{kenn}. This ratio of luminosities represents the fraction of unobscured star formation. The factor $\alpha$ in the denominator is the proportionality factor adopted from \citet{2008ApJ...686..155Z}. For $\alpha$ = 6.31, they found that the sum $L_{\text{FUV,obs}}$ + 6.31 $L_{24\mu\text{m}}$ is a good approximation of the attenuation corrected FUV flux.}
  \label{plot_sfrzhu_ratiofuv24.gif}
\end{figure}

Relying on the relation in \citet{2008ApJ...686..155Z}, we can quantify the fraction of dust-enshrouded and unobscured star formation. 
Figure \ref{plot_sfrzhu_ratiofuv24.gif} plots the ratio between the observed FUV luminosity and the scaled MIPS  24$\mu$m luminosity, for which the scaling coefficient $\alpha$ = 6.31 was adopted from \citet{2008ApJ...686..155Z}, as a function of the SFR. This $L_{\text{FUV}}$-to-6.31$L_{\text{24}\mu\text{m}}$ ratio represents the unobscured fraction of star formation. Figure \ref{plot_sfrzhu_ratiofuv24.gif} illustrates that the majority of our sample galaxies have a dominant contribution to the 24$\mu$m emission from dust grains heated by a young stellar population, rather than direct FUV emission from young stars. This makes the 24$\mu$m or a combination of GALEX FUV data and an appropriate attenuation correction a more reliable star formation indicator for our sample than solely relying on UV data. One exception is NGC\,4278, for which the unobscured fraction of star formation dominates dust-enshrouded 24$\mu$m emission of young stellar objects. Furthermore, Figure \ref{plot_sfrzhu_ratiofuv24.gif} shows a hint for a decreasing trend in the unobscured fraction of star formation towards higher SFR. This trend can be accounted for by an increasing opacity for galaxies which form stars at a higher rate.

\begin{table}
\centering
\caption{Star formation rates for the galaxies in our sample, derived from different SFR relations based on MIPS 24$\mu$m data, or a combination of GALEX FUV with MIPS 24$\mu$m data.}
\begin{tabular}{lcc}
  \hline\hline \\
  Name & Rieke2009 &  Zhu2008+Kennicutt2009 \\ 
            & (M$_{\text{\sun}}$ yr$^{-1}$) & (M$_{\text{\sun}}$ yr$^{-1}$) \\
 (1) & (2) & (3)\\
 \\ \hline \\
 Cartwheel     & 3.25 $\pm$ 1.14 &  7.97 $\pm$ 1.63  \\
 NGC\,0520   & 5.04 $\pm$ 1.77& 7.02 $\pm$ 1.43  \\
 NGC\,0625  & 0.04 $\pm$ 0.01&  0.06 $\pm$ 0.01 \\    
 NGC\,0660   & 1.23 $\pm$ 0.43 & 1.70  $\pm$ 0.35 \\    
 NGC\,0695   & 42.37 $\pm$ 14.88 & 62.63 $\pm$ 12.78  \\    
 NGC\,0986   & 3.87 $\pm$ 1.36&  5.39 $\pm$ 1.10 	 \\    
 UGC\,02238 &13.04 $\pm$ 4.58 & 18.19 $\pm$ 3.72   \\    
 NGC\,1156   & 0.06 $\pm$ 0.02 & 0.18 $\pm$ 0.04  \\    
 NGC\,1266  & 2.40 $\pm$ 0.84 & 3.34 $\pm$ 0.68  \\    
 NGC\,1317   & 0.18 $\pm$ 0.06 &  0.31 $\pm$ 0.06  \\      
 NGC\,2388   & 17.31 $\pm$ 6.08 &  23.35 $\pm$ 4.79  \\ 
 NGC\,4041   & 1.82 $\pm$ 0.64 &  2.62 $\pm$ 0.53 \\ 
 NGC\,4189   & 0.25 $\pm$ 0.09 &  0.42 $\pm$ 0.09  \\ 
 NGC\,4278   & 0.014 $\pm$ 0.005 &  0.03 $\pm$ 0.01 \\ 
 NGC\,4293   & 0.46 $\pm$ 0.16 &  0.64 $\pm$  0.13  \\ 
 NGC\,4299   & 0.18 $\pm$ 0.06 & 0.52 $\pm$ 0.11  \\ 
 NGC\,4490   & 0.26 $\pm$ 0.09 & 0.42 $\pm$ 0.08 \\ 
 NGC\,4651   & 0.47 $\pm$ 0.17 &  0.87 $\pm$ 0.18  \\  
 NGC\,5713   & 6.04 $\pm$ 2.12 &  8.73 $\pm$ 1.78 \\
 NGC\,6240   & 101.85 $\pm$ 35.78 & 127.15 $\pm$ 25.94  \\
 NGC\,7217   & 0.30 $\pm$ 0.11 &  0.49 $\pm$ 0.10 \\
 NGC\,7552   & 10.99 $\pm$ 3.86 & 15.26 $\pm$ 3.11  \\
 NGC\,7714   & 8.85 $\pm$ 3.11 & 13.34 $\pm$ 2.72  \\
 NGC\,7771   & 14.06 $\pm$ 4.94 & 19.47 $\pm$ 3.97  \\
  \hline\hline
\end{tabular}
\label{Table3}
\end{table}

\begin{table*}
\centering
\caption{Coefficients and uncertainties of the best fitting line for the relation between $L_{[\text{C{\sc{ii}}}]}$ and two SFR relations.}
\begin{tabular}{lccc}
  \hline\hline \\
Variable & Slope & Intercept & 1$\sigma$ dispersion (dex)\\   
 \\ \hline \\ 
\\        
SFR$_{24\mu m,Rieke2009}$ & 1.040$\pm$0.035  & -42.535$\pm$1.441 & 0.30\\
\\
SFR$_{FUV+24\mu m,Zhu2008+Kennicutt2009}$ & 0.983$\pm$0.021  & -40.012$\pm$0.858  & 0.27\\
\\
  \hline\hline
\end{tabular}
\label{Table4}
\end{table*}

\section{Discussion}

\subsection{[C{\sc{ii}}] as a SFR indicator}

In the previous section, we found that the star formation rate correlates well with the [C{\sc{ii}}] luminosity. Table \ref{Table3} summarizes the star formation rates obtained from the two different star formation relations for all galaxies in our reduced sample. Table \ref{Table4} summarizes the coefficients $a$ and $b$ for the best fitting line $y = ax+b$ for each star formation tracer, the uncertainty on the slope $a$ and intercept $b$ and the 1$\sigma$ dispersion of the individual galaxies around this mean trend. 
For both star formation relations, the uncertainties on the slope and intercept of the best fitting line are small and the spread around the mean trend is narrow. The dispersion around the correlation is smaller for the SFR derived from a combination of GALEX FUV and 24$\mu$m data (0.27 dex), than when estimating the SFR from the single 24$\mu$m luminosities (0.30 dex) (see also the bottom panels of Figures \ref{plot_sfrrieke_lcii.gif} and \ref{plot_sfrzhu_lcii.gif}). 
Considering that a combination of FUV and 24$\mu$m data traces the complete star formation activity, as opposed to the single 24$\mu$m data, which only traces the obscured fraction of star formation, we will use the SFR tracing both the dust-enshrouded and unobscured activity to calibrate the SFR relation. Since the dispersion is smaller for the SFR when estimated from both the FUV and 24$\mu$m luminosity (0.27) rather than from the 24$\mu$m data (0.30 dex), we believe this as a confirmation of the link between the [C{\sc{ii}}] emission and star formation activity in a galaxy, that has been reported in this work.

From Eq. \ref{FUV+24} we derive the SFR calibration:
\begin{equation}
\label{eqcali}
\text{SFR} = \frac{(L_{[\text{C{\sc{ii}}}]})^{0.983}}{1.028 \times 10^{40}} 
\end{equation}
where the SFR and $L_{[\text{C{\sc{ii}}}]}$ are in units of M$_{\odot}$ yr$^{-1}$ and erg s$^{-1}$, respectively. The SFR calibration factors are derived assuming a \citet{2001MNRAS.322..231K} IMF. This relation is valid for star-forming, late-type galaxies with a star formation activity in the range  0.03 - 127 M$_{\text{\sun}}$ yr$^{-1}$ and objects with a [C{\sc{ii}}] luminosity between 39.3 $\la$ $\log$ $L_{[\text{C{\sc{ii}}}]}$[erg s$^{-1}$] $\la$ 42.7 and a TIR luminosity between 42.1 $\la$ log $L_{\text{TIR}}$[erg s$^{-1}$] $\la$ 45.4.

\begin{figure}
\centering
\includegraphics[width=0.45\textwidth]{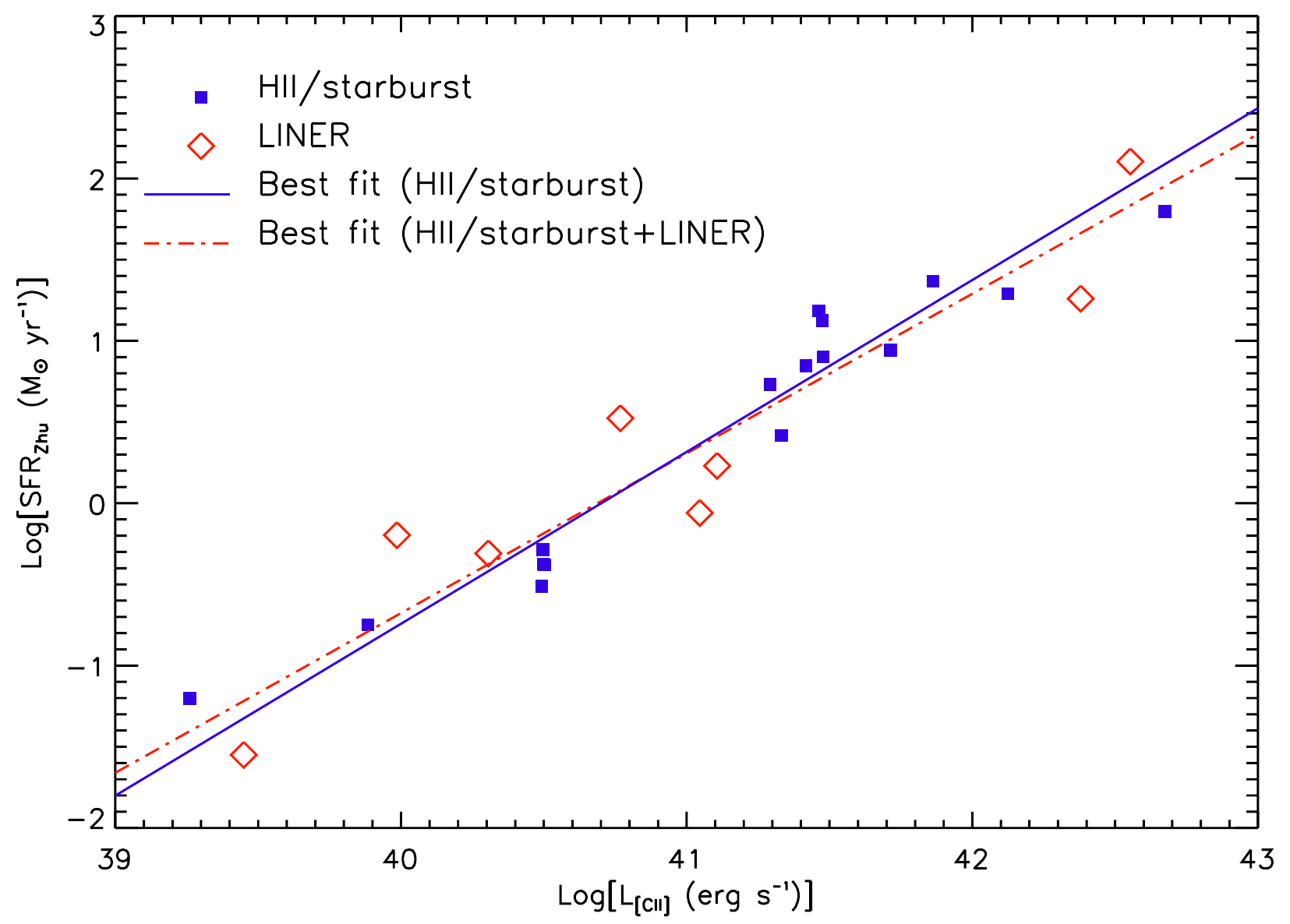}
 \caption{
SFR$_{FUV+24\mu m}$ (attenuation correction and SFR relation from \citealt{2008ApJ...686..155Z} and \citealt{2009ApJ...703.1672K}, respectively) as a function of $L_{[\text{C{\sc{ii}}}]}$. In this plot, H{\sc{ii}}/starburst and LINER galaxies are indicated as blue filled squares and red diamonds, respectively. The best fitting line for the H{\sc{ii}}/starburst sample and the combined H{\sc{ii}}/starburst+LINER sample are shown as blue plain and red dashed-dotted lines, respectively.}
 \label{plot_hii_liner.gif}
\end{figure}

\begin{figure}
\centering
\includegraphics[width=0.45\textwidth]{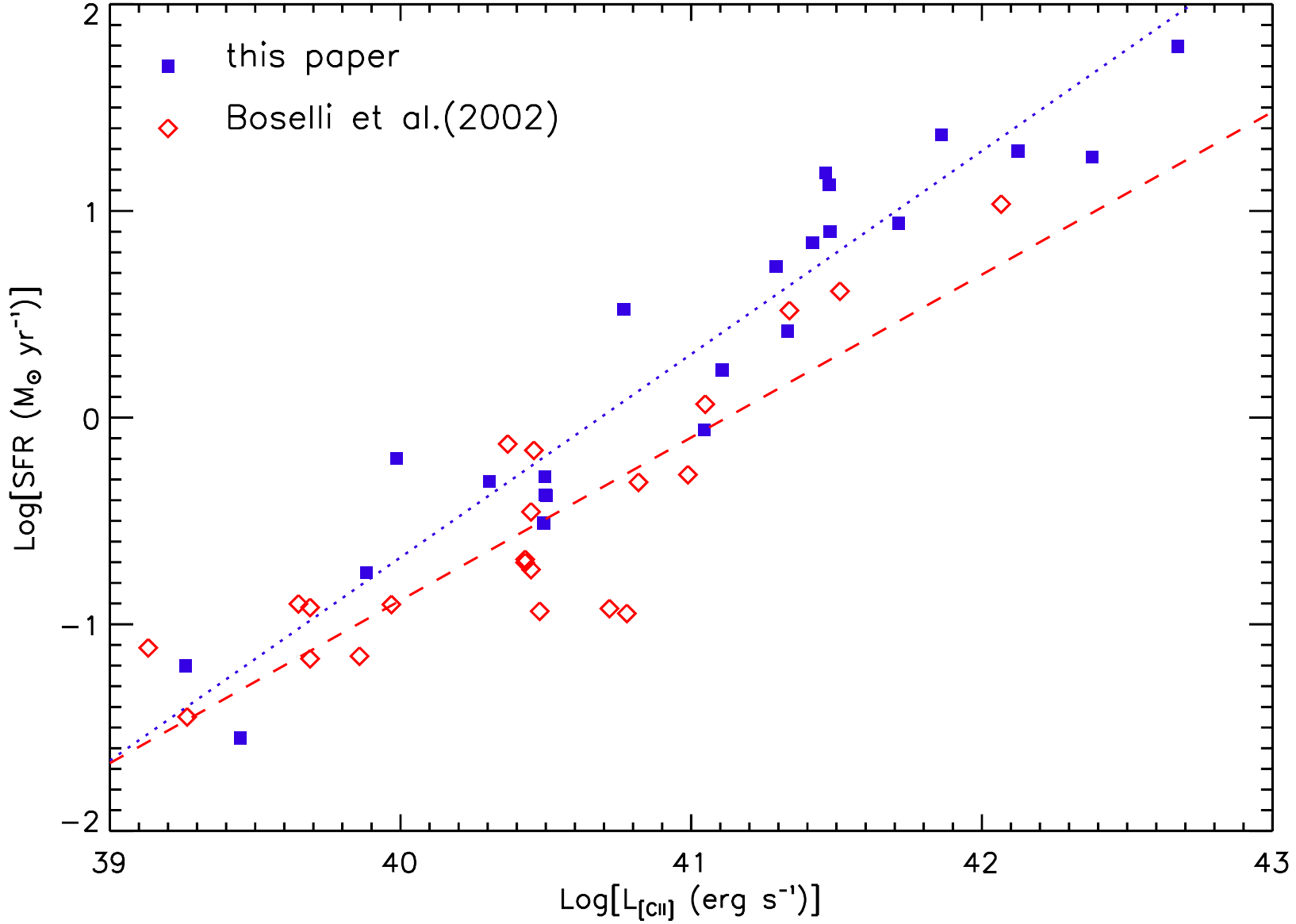}
 \caption{
The star formation rate as estimated from relation \ref{eqcali} in this paper (blue filled squares) and from the SFR calibration in \citet{2002A&A...385..454B}. Since \citet{2002A&A...385..454B} assumed a \citet{1955ApJ...121..161S} IMF ($\alpha$ = 2.35) in the mass range between 0.1 and 100 M$_{\odot}$, we have divided their calibration coefficient by 1.51, to convert it to the \citet{2001MNRAS.322..231K} IMF we applied. The best fitting lines for our calibration and the one in \citet{2002A&A...385..454B} are indicated as blue dotted and red dashed lines, respectively.}
 \label{plot_compare.gif}
\end{figure}

Since the residual plot of this SFR relation (see Figure \ref{plot_sfrzhu_lcii.gif}, bottom) indicates that the LINER galaxies in are sample contribute most to the observed scatter, we investigate whether a possible AGN contribution to the LINER galaxies introduce a bias to the derived SFR relation.
If we calibrate the SFR relation for the 16 H{\sc{ii}} and starburst galaxies in our sample, we should see a significant difference if the LINER sample indeed introduces a bias, possibly due to a small AGN contribution to the 24$\mu$m luminosities. 
From Figure \ref{plot_hii_liner.gif} we can deduce that the best fitting line in our sample does not change substantially if we neglect the LINER galaxies for the SFR calibration. This already shows that the LINER sample does not introduce any bias due to a small AGN contribution. Moreover, half of the LINER galaxies are situated below the best fitting line.

From this tight correlation between the star formation rate and the [C{\sc{ii}}] luminosity, we conclude that the [C{\sc{ii}}] luminosity is a good star formation rate indicator.  
The applicability of [C{\sc{ii}}] as a star formation rate diagnostic had already been hinted by several authors \citep{1991ApJ...373..423S,1999MNRAS.303L..29P,1999MNRAS.310..317L,2002A&A...385..454B,2010ApJ...724..957S}, but we were able to quantify this correlation in an accurate way. 
Figure \ref{plot_compare.gif} compares our SFR relation to the calibration obtained in \citet{2002A&A...385..454B}. Since \citet{2002A&A...385..454B} assumed a \citet{1955ApJ...121..161S} IMF ($\alpha$ = -2.35) in the mass range between 0.1 and 100 M$_{\odot}$, we have divided their calibration coefficient by 1.51, to convert it to the \citet{2001MNRAS.322..231K} IMF (see \citet{2010ApJ...714.1256C} for the derivation of the factor 1.51 difference between the \citet{1955ApJ...121..161S} and \citet{2001MNRAS.322..231K} IMF calibrations). For low [C{\sc{ii}}] luminosity objects both calibrations are still quite consistent, but for galaxies with an increasing [C{\sc{ii}}] luminosity the SFR estimate provided by \citet{2002A&A...385..454B} quickly diverges from our estimates up to a factor of $\sim$5 for the highest luminosity objects in our sample. This deviation towards higher [C{\sc{ii}}] luminosities is probably due to an underestimation of the H$\alpha$ extinction in more luminous galaxies, since the attenuation relations at that time where calibrated without taking into account heavily obscured star formation and thus underestimating the true H$\alpha$ emission. The scatter around the mean trend in our SFR relation is smaller than a factor of 2, while the dispersion in the [C{\sc{ii}}]-H$_{\alpha}$ luminosity relation already reaches a factor of $\sim$4. The final uncertainty on the SFR calibration in \citet{2002A&A...385..454B} is estimated to be as high as a factor of $\sim$ 10. We believe the poorly known characteristics of the [N{\sc{ii}}] contamination and the attenuation correction at the time \citet{2002A&A...385..454B} performed their SFR calibration explain a significant amount of their reported scatter. Benefiting from the increased database of [C{\sc{ii}}] observations and exploring other reliable SFR tracers in this paper, we can revise the analysis in \citet{2002A&A...385..454B} and conclude that the [C{\sc{ii}}] emission is a reliable SFR indicator in most normal star-forming galaxies.

\subsection{Applicability}
\label{Applicability}
\begin{figure}
\centering
\includegraphics[width=0.45\textwidth]{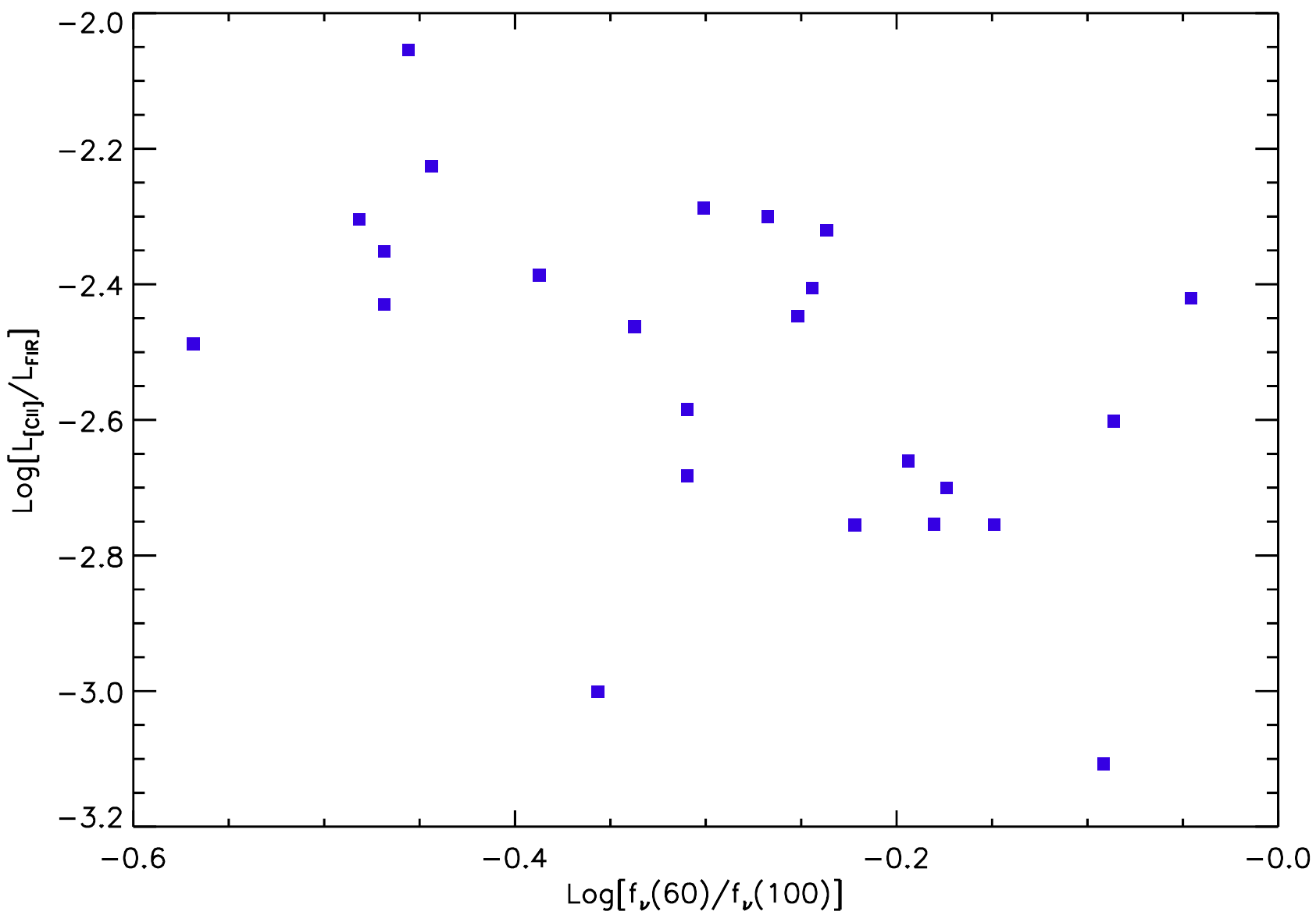}
 \caption{The ratio $L_{[\text{C{\sc{ii}}}]}$-to-$L_{\text{FIR}}$ as a function of the effective dust temperature, as embodied by the $f_{\nu}$(60$\mu$m)-to-$f_{\nu}$(100$\mu$m) ratio, where the FIR luminosity was derived following \citet{1985ApJ...298L...7H}. We notice a decrease of the $L_{[\text{C{\sc{ii}}}]}$-to-
$L_{\text{FIR}}$ ratio with increasing warm dust temperature.}
 \label{plot_ratiociifir_ratio60100.gif}
\end{figure}

The extension of this SFR relation to ULIRGS should be treated with caution, since a decrease of the $L_{[\text{C{\sc{ii}}}]}$-to-$L_{\text{FIR}}$ ratio with increasing warm infrared color for all galaxy types has been observed in many samples \citep{1985ApJ...291..755C,1991ApJ...373..423S,1997ApJ...491L..27M,2001ApJ...561..766M,2003ApJ...594..758L,2005SSRv..119..355V,2008ApJS..178..280B}. More specifically, \citet{2003ApJ...594..758L} report a [C{\sc{ii}}] line deficit in a sample of 15 ULIRGS, after they had noticed this trend in a few individual ultra-luminous objects \citep{1998ApJ...504L..11L}. Figure \ref{plot_ratiociifir_ratio60100.gif} hints at a similar, but weaker, trend for the galaxies in our sample. This effect is probably due to a decrease in efficiency of the photoelectric heating of the gas in strong radiation fields \citep{2001ApJ...561..766M}. Other possible explanations for this trend were given in \citet{2001A&A...375..566N} and \citet{2000A&A...359...41B}. \citet{2001A&A...375..566N} suggest an increased collisional de-excitation of [C{\sc{ii}}] due to an enhanced gas density, or a decrease in the ionized component for increasing star formation activity. While \citet{2000A&A...359...41B} invokes the self-absorption of [C{\sc{ii}}] for galaxies with increasing metallicity. 
Recently, \citet{2010ApJ...711..757P} and \citet{Rangwala} claimed that a high dust optical depth could be the cause of this [C{\sc{ii}}] line deficit in several galaxies.
If this trend in the [C{\sc{ii}}] line deficit holds for more IR-luminous galaxies, this might have its implications for the reliability of [C{\sc{ii}}] as a star formation indicator in those objects. 
Even more recently, \citet{2011ApJ...728L...7G} inferred from their Herschel observations a similar line deficit for several other FIR fine structure lines ([N{\sc{ii}}], [O{\sc{i}}] , [N{\sc{iii}}] , [O{\sc{iii}}]). They allocate this deficit to a transition between two modes of star formation (in normal disk galaxies and major merger systems) with a different star formation efficiency.

On the other hand, \citet{1999MNRAS.303L..29P} also report a decrease in the $L_{[\text{C{\sc{ii}}}]}$-to-$L_{\text{FIR}}$ ratio for decreasing H$\alpha$ EW. This trend has been observed for more quiescent galaxies with H$\alpha$ EW $\le$ 10 \AA~and can be explained by an increasing contribution from older, less massive stars that heat dust grains and therefore contribute to the FIR-luminosity. The radiation from these stars is not hard enough to substantially heat the dust grains through the photoelectric effect \citep{2003A&A...397..871P}. Since collisions with these photoelectrons are the main heating source for the interstellar gas, this inefficient gas heating will render those galaxies [C{\sc{ii}}] quiet for an increasing contribution from a more evolved stellar population. This effect makes the extension from the SFR relation to quiescent galaxies not straightforward either. For objects with an extremely low star formation activity, \citet{1999MNRAS.303L..29P} also found that the main contribution to the [C{\sc{ii}}] emission arises from the diffuse neutral interstellar medium. In our current sample, two galaxies (NGC\,4293 and NGC\,7217) satisfy the criterion H$\alpha$ EW $\le$ 10 \AA,~characterizing a quiescent galaxy \citep{1983AJ.....88.1094K,2008ApJS..178..247K}. Since those galaxies do not deviate from the mean trend in the SFR-$L_{[\text{C{\sc{ii}}}]}$ plot, a larger sample of quiescent galaxies is necessary to further investigate whether our SFR relation is expandable to objects with a lower SFR compared to the galaxies in the present sample. 

Because of systematic variations in the $L_{[\text{C{\sc{ii}}}]}$-to-$L_{\text{FIR}}$ ratio and varying contribution with luminosity from several components (PDRs, diffuse cold neutral and warm ionized gas, H{\sc{ii}} regions) to the [C{\sc{ii}}] emission, we should be careful in extending the SFR relation in this paper to more extreme luminosities, on both ends of the luminosity scale. Although the derived star formation relation might not be valid for more extreme cases of star formation activity (either very quiescent or actively star-forming), this calibration might be useful as an alternative indicator of the star formation rate in objects which lack either UV or IR observations. Since [C{\sc{ii}}] emission is hardly not affected by attenuation in most cases (currently, Arp 220 is the only object where prominent obscuration effects have been claimed), it might provide an immediate probe of the complete star formation activity. 

At high redshift, the behaviour of the $L_{[\text{C{\sc{ii}}}]}$-to-$L_{\text{FIR}}$ ratio is even different. Ultra luminous objects (ULIRGs; $L_{\text{TIR}}$ $>$ 10$^{12}$ $L_{\odot}$) do not seem te be affected by a [C{\sc{ii}}] deficit. On the contrary, \citet{2009A&A...500L...1M} suggest an enhancement by at least one order of magnitude of the [C{\sc{ii}}] emission in galaxies at high-redshift compared to local galaxies of the same infrared luminosity. \citet{2009A&A...500L...1M} argues that this enhanced [C{\sc{ii}}] emission could be due to lower metallicities of the ISM in those high-redshift galaxies, which tend to have a lower dust content and therefore a larger [C{\sc{ii}}] emitting region. Independent of the physical origin of this effect, if confirmed by future observations, the strong [C{\sc{ii}}] emission from IR luminous high-redshift objects would have great implications for future observations at high redshift. This suggested increase in detectability of the [C{\sc{ii}}] line at high redshift would also imply an extensive applicability of a SFR relation based on [C{\sc{ii}}] luminosities, which we will derive in this paper.

Recently, \citet{2010ApJ...724..957S} reported the detection of the [C{\sc{ii}}] line in 12 galaxies at redshifts ranging from 1 to 2. They concluded that for starburst-powered galaxies the [C{\sc{ii}}] emission is comparable to values found in local star-forming galaxies with similar FIR luminosities ([C{\sc{ii}}]/FIR $\sim$ 3$\times$10$^{-3}$), while AGN-dominated galaxies are characterized by [C{\sc{ii}}]-to-FIR luminosity ratios similar to local ULIRGs ([C{\sc{ii}}]/FIR $\sim$ 4$\times$10$^{-4}$), suggesting that this effect of enhanced [C{\sc{ii}}] emission only becomes apparent at even higher redshift (z$>$2). 
\citet{2010ApJ...724..957S} also reported that [C{\sc{ii}}] is a reliable star formation indicator for their heterogeneous sample of starburst- and/or AGN-dominated galaxies in the  redshift interval z = [1,2]. At even higher redshifts (z$>$2), the lower metallicity might influence the temperature and chemical structure of PDRs and thus, the [C{\sc{ii}}] emission (e.g. \citealt{1994A&A...292..371L,1995ApJ...443..152W,2006A&A...451..917R}). Future observations of ULIRGs with Herschel and ALMA will provide insight in the applicability of [C{\sc{ii}}] as a star formation indicator at high redshift and the validity of the star formation relation derived in this paper. 

\subsection{Nature of the [C{\sc{ii}}] emission in galaxies}

Besides the first accurate quantitative calibration of the star formation rate against the [C{\sc{ii}}] luminosity, the tightness of this correlation gives us insight in the origin of the [C{\sc{ii}}] emission on a global galaxy-scale. In this section, we introduce two possible explanations for the tightness of the SFR-[C{\sc{ii}}] relation.

Although PDRs are the main contributor to the [C{\sc{ii}}] emission in most galaxies, a significant fraction also originates in the cold neutral medium (CNM) (i.e. H{\sc{i}} clouds, \citet{1987ASSL..134...87K, 1993ApJ...407..579M,1994ApJ...434..587B,1995ApJ...443..152W, 1998A&A...339...19S,1999MNRAS.303L..29P,2001A&A...373..827P,2002AJ....124..751C}), the warm ionized medium (WIM) (i.e. diffuse H{\sc{ii}} regions, \citet{1993ApJ...407..579M,1994ApJ...436..720H,1997ApJ...491L..27M,2001ApJ...561..766M,1999MNRAS.310..317L,2002AJ....124..751C}) and to lesser extend also in H{\sc{ii}} regions \citep{1990A&AS...83..501S} (see \citealt{2002A&A...385..454B} for a quantitative analysis of these different contributions). 
Since the contribution from compact H{\sc{ii}} regions to the [C{\sc{ii}}] emission is negligible with respect to other components on galactic scales (e.g. \citealt{1990A&AS...83..501S,2000ApJ...543..634M}), it is difficult to believe that the contribution from these star-forming regions alone causes this good correlation. 

This brings us to a first possible explanation for the strong correlation between [C{\sc{ii}}]  and the SFR.
Considering that PDRs are neutral regions of warm dense gas at the boundaries between H{\sc{ii}} regions and molecular clouds, we think that most of the [C{\sc{ii}}] emission from PDRs arises from the immediate surroundings of star-forming regions. This might not be surprising, since FUV photons from young O and B stars escaping from the dense H{\sc{ii}} regions, impinge on the surface of these neutral PDR regions where they will heat the gas through the photoelectric effect on dust grains.
We believe that a more or less constant contribution from PDRs to the [C{\sc{ii}}] emission and the fact that this [C{\sc{ii}}] emission from PDRs stems from the outer layers of photon-dominated molecular clumps (i.e. at the boundary of molecular clouds and H{\sc{ii}} regions) might be a reasonable explanation for the tight correlation between the star formation rate and $L_{[\text{C{\sc{ii}}}]}$. 
Also, \citet{2001ApJ...561..766M} suggest this tight correlation between PDRs and star-forming regions from their analysis of the FUV flux $G_{0}$ and gas density $n$ in PDRs. The high PDR temperature and pressure required to fit their data, imply that most of the line and continuum FIR emission arises from the immediate proximity of expanding H{\sc{ii}} regions.
Moreover, mapping of [C{\sc{ii}}] in the Milky Way \citep{1994ApJ...434..587B} and in spatially resolved nearby galaxies (see \citealt{2005SSRv..119..313S} for an overview) gives indications for a close association between PDRs and ionized gas in these galaxies.

Alternatively, we consider the likely possibility that this tight correlation is not the reflection of shared photoexcitation processes, taking place at the same position within a galaxy. 
On the contrary, the [C{\sc{ii}}] emission might not be directly linked to the star formation, but could instead trace the cold ISM being therefore indirectly related to the star formation through the Schmidt law \citep{1998ApJ...498..541K}. 
Evidence confirming this indirect link is the association of [C{\sc{ii}}] emission with modest densities and softer FUV radiation fields (e.g \citealt{2010A&A...521L..19P}).

A similar connection between the [C{\sc{ii}}] emission in a galaxy and the diffuse ISM is typically present in more quiescent objects \citep{1999MNRAS.303L..29P, 2001A&A...373..827P, 2003A&A...397..871P}, while in star forming galaxies the [C{\sc{ii}}] line emission is found to mainly arise from PDR's \citep{1985ApJ...291..755C,1991ApJ...373..423S,1993ApJ...407..579M}. Mapping of [C{\sc{ii}}] at high resolution is necessary to ascertain the nature of the tight correlation between the SFR and [C{\sc{ii}}] emission for the normal star-forming galaxies in our sample.

Future observations of spatially resolved objects in the nearby universe and at high redshift of [C{\sc{ii}}], CO(1-0) and [N{\sc{ii}}] lines for a sample of galaxies ranging several orders of magnitude in IR-luminosity, will be able to disentangle the different components contributing to the [C{\sc{ii}}] emission, since the CO(1-0) intensity correlates well with [C{\sc{ii}}] \citep{1999RvMP...71..173H} in PDRs and [N{\sc{ii}}] traces the H{\sc{ii}} regions and the warm ionized medium \citep{1991ApJ...381..200W, 1994ApJ...434..587B}. Such analysis will enable us to make a distinction between the different sources that contribute to the [C{\sc{ii}}] emission and, in particular, examine the tight correlation between the star formation rate and [C{\sc{ii}}] luminosity. 
High spatial resolution data will be able to distinguish whether [C{\sc{ii}}] emission from PDRs is closely related to star-forming regions in galaxies or rather the global [C{\sc{ii}}] emission in a galaxy traces the gas mass. The latter would imply an indirect link between [C{\sc{ii}}]  and the star formation activity in a galaxy through the Schmidt law.

\section{Conclusions}
The aim of this paper was to investigate the reliability of [C{\sc{ii}}] as a SFR indicator. The analysis was conducted in the following way:
\begin{itemize}
\item We assembled a sample of 38 galaxies for which ISO [C{\sc{ii}}], GALEX FUV and MIPS 24 $\mu$m data are available. This initial sample was subdivided according to their nuclear spectral classification (H{\sc{ii}}/starburst, LINER, AGN). Due to the contaminating 24$\mu$m emission in AGN hosts, we were unable to directly relate the 24$\mu$m emission in those objects to the star formation activity. Therefore, the objects with nuclear spectra resembling those of an AGN, were eliminated from our sample.
\item For the remaining 24 galaxies in our sample, the SFR was estimated from either the 24$\mu$m luminosity or a combination of FUV and 24$\mu$m data. In the latter case the FUV emission is corrected for internal dust attenuation according to \citet{2008ApJ...686..155Z}. The SFR calibration in \citet{2009ApJ...703.1672K} provides a SFR estimate for this extinction-corrected FUV luminosity. 
Comparing the dispersions for these two different SFR relations, we found that the SFR correlates best with the [C{\sc{ii}}] luminosity, when the star formation activity is traced by a combination of those luminosities.
\item From the tight correlation of the SFR with the [C{\sc{ii}}] luminosity (the dispersion around the mean trend is 0.27 dex), we conclude that the [C{\sc{ii}}] luminosity is a reliable SFR diagnostic in normal, star-forming galaxies in the local universe and find the following SFR calibration:
\begin{equation}
\label{eqcali}
\text{SFR} [\text{M}_{\odot} \text{yr}^{-1}] = \frac{(L_{[\text{C{\sc{ii}}}]}[\text{erg s}^{-1}])^{0.983}}{1.028 \times 10^{40}}
\end{equation}
\item The extension of this relation to more quiescent galaxies (H$\alpha$ EW $\le$ 10 \AA) and ultra luminous galaxies (ULIRGs) should be treated with caution, due to a deviation from the $L_{[\text{C{\sc{ii}}}]}$-to-$L_{\text{FIR}}$ correlation at those extreme ends of the luminosity scale.  
\item For galaxies in the redshift range z=[1,2], the comparable [C{\sc{ii}}] luminosities and the enhanced [C{\sc{ii}}] emission, at even higher redshift, compared to local galaxies with similar FIR luminosities, will render an immense applicability of this SFR relation in future high redshift surveys (both with Herschel and ALMA).
\item We believe this tight correlation between the [C{\sc{ii}}] luminosity and the star formation activity may imply a direct link in photo-excitation processes: the [C{\sc{ii}}] emission from PDRs arises from the immediate surroundings of star-forming regions (i.e. from the outer layers of photon-dominated clumps) and contributes a more or less constant fraction on a global galaxy-scale. Alternatively, the [C{\sc{ii}}] emission might equally well trace the star formation in a galaxy, when assuming that the [C{\sc{ii}}] emission is associated to the cold ISM in a galaxy and, therefore, is indirectly linked to the SFR through the Schmidt law $\Sigma_{SFR}$ = $\Sigma_{gas}^{1.4}$.
\end{itemize}

\section*{Acknowledgements}

We thank the referee for his/her comments, which helped us to improve the paper considerably.

GALEX is a NASA Small Explorer, launched in 2003 April. We gratefully acknowledge NASA's support for construction, operation and science analysis for the GALEX mission, developed in cooperation with the Centre National d'Etudes Spatiales (CNES) of France and the Korean Ministry of Science and
Technology.

This work is based in part on observations made with the Spitzer Space Telescope, which is operated by the Jet Propulsion Laboratory, California Institute of Technology under a contract with NASA.

\appendix

\bsp

\label{lastpage}

\end{document}